\documentclass[aps,prd,12pt,notitlepage,showpacs,nofootinbib,tightenlines]{revtex4-1}
\usepackage{amsmath}
\usepackage{amssymb}
\usepackage{epsfig}
\usepackage{graphicx}
\usepackage{bm}
\usepackage{times}
\usepackage{braket}
\usepackage{color}
\usepackage{slashed}
\usepackage{hyperref}
\definecolor{Red}{rgb}{1.,0.,0.}

\definecolor{Blue}{rgb}{0.,0.,1.}

\definecolor{nicered}{rgb}{0.7,0.1,0.1}
\definecolor{nicegreen}{rgb}{0.1,0.5,0.1}
\hypersetup{colorlinks,citecolor=nicegreen,linkcolor=nicered}

\begin{document}

\newcommand{\beq}{\begin{eqnarray}}
\newcommand{\eeq}{\end{eqnarray}}
\newcommand{\ben}{\begin{enumerate}}
\newcommand{\een}{\end{enumerate}}
\newcommand{\non}{\nonumber\\ }

\newcommand{\jpsi}{J/\Psi}

\newcommand{\ppa}{\phi_\pi^{\rm A}}
\newcommand{\ppp}{\phi_\pi^{\rm P}}
\newcommand{\ppt}{\phi_\pi^{\rm T}}
\newcommand{\ov}{ \overline }

\newcommand{\zerot}{ {\textbf 0_{\rm T}} }
\newcommand{\kt}{k_{\rm T} }
\newcommand{\fb}{f_{\rm B} }
\newcommand{\fk}{f_{\rm K} }
\newcommand{\rk}{r_{\rm K} }
\newcommand{\mb}{m_{\rm B} }
\newcommand{\mw}{m_{\rm W} }
\newcommand{\im}{{\rm Im} }

\newcommand{\kks}{K^{(*)}}
\newcommand{\acp}{{\cal A}_{\rm CP}}
\newcommand{\pb}{\phi_{\rm B}}

\newcommand{\xeba}{\bar{x}_2}
\newcommand{\xsba}{\bar{x}_3}
\newcommand{\peas}{\phi^A}

\newcommand{\pvsl}{ p \hspace{-2.0truemm}/_{K^*} }
\newcommand{\esl}{ \epsilon \hspace{-2.1truemm}/ }
\newcommand{\psl}{ p \hspace{-2truemm}/ }
\newcommand{\ksl}{ k \hspace{-2.2truemm}/ }
\newcommand{\lsl}{ l \hspace{-2.2truemm}/ }
\newcommand{\nsl}{ n \hspace{-2.2truemm}/ }
\newcommand{\vsl}{ v \hspace{-2.2truemm}/ }
\newcommand{\epsl}{\epsilon \hspace{-1.8truemm}/\,  }
\newcommand{\bfkk}{{\bf k} }
\newcommand{\calm}{ {\cal M} }
\newcommand{\calh}{ {\cal H} }
\newcommand{\calo}{ {\cal O} }

\def \appb{{\bf Acta. Phys. Polon. B }  }
\def \cpc{ {\bf Chin. Phys. C } }
\def \ctp{ {\bf Commun. Theor. Phys. } }
\def \epjc{{\bf Eur. Phys. J. C} }
\def \ijmpcs{{\bf Int. J. Mod. Phys. Conf. Ser.} }
\def \jhep{{\bf J. High Energy Phys. } }
\def \jpg{ {\bf J. Phys. G} }
\def \mpla{{\bf Mod. Phys. Lett. A } }
\def \npb{ {\bf Nucl. Phys. B} }
\def \plb{ {\bf Phys. Lett. B} }
\def \ppn{ {\bf Phys. Part. Nucl. } }
\def \ppnp{{\bf Prog.Part. Nucl. Phys.  } }
\def \pr{  {\bf Phys. Rep.} }
\def \prc{ {\bf Phys. Rev. C }}
\def \prd{ {\bf Phys. Rev. D} }
\def \prl{ {\bf Phys. Rev. Lett.}  }
\def \ptp{ {\bf Prog. Theor. Phys. }}
\def \zpc{ {\bf Z. Phys. C}  }
\def \jpg{ {\bf J.Phys.-G-}  }
\def \ap{ {\bf Ann. of Phys}  }

\title{The perturbative QCD factorization of $\rho \gamma^{\star} \to \rho$}
\author{Ya-lan Zhang$^{1}$}
\author{Shan Cheng$^{2}$} \email{cheng@physik.uni-siegen.de}
\author{Jun Hua$^{1}$}
\author{Zhen-Jun Xiao$^{1,3}$ } \email{xiaozhenjun@njnu.edu.cn}
\affiliation{1.  Department of Physics and Institute of Theoretical Physics,
Nanjing Normal University, Nanjing, Jiangsu 210023, People's Republic of China,}
\affiliation{2.  Theoretische Elementarteilchenphysik, Naturwissenschaftlich Technische Fakult$\ddot{a}$t,
Universi$\ddot{a}$t Siegen, 57068 Siegen, Germany,}
\affiliation{3. Jiangsu Key Laboratory for Numerical Simulation of Large Scale Complex Systems,
Nanjing Normal University, Nanjing 210023, People's Republic of China}
\date{\today}
\begin{abstract}
In this paper we firstly demonstrate step by step that the factorization hypothesis is valid at
the next-to-leading order (NLO)
for the exclusive process $\rho \gamma^{\star} \to \rho$ by employing the collinear factorization
approach, and then extend this proof to the case of the $\kt$ factorization by taking into account the transversal
momentum of the light external quark (anti-quark) lines in the $\rho$ meson.
At the NLO level, we then show that the soft divergences from different sub-diagrams will be canceled
each other in the quark level,
while the remaining collinear divergences can be absorbed into the NLO meson wave functions.
The full NLO amplitudes can therefore be factorized as the convolution of the NLO wave functions $ \Phi^{(1)}_{\rho}$
and the infrared-finite leading order (LO) hard kernels  $G^0_{X,IJ,kl}$ in the $\kt$ factorization.
We also write down the polarized NLO $\rho$ meson wave functions  in the form of nonlocal hadron
matrix elements with the gauge factor integral path deviating from the light cone.
These NLO $\rho$ meson wave functions can be used to calculate the NLO hard corrections to
some relevant exclusive processes, such as $B \to \rho$ transition.
\end{abstract}



\maketitle

\section{Introduction}

The factorization theorem \cite{plb87-359,prd55-272,prp112-173} is a fundamental hypothesis to deal with
the QCD involved exclusive processes with various energy scales which may generate infrared
divergence \cite{end-div}.
The perturbative QCD (PQCD) factorization approach is constructed based on the so-called
$\kt$ factorization theorem \cite{kt-fact}, where the parton transversal momentum in the denominator of the
propagators are picked up in order to remove the end-point singularity.
In the PQCD factorization approach, fortunately, the hard contributions can be calculated
perturbatively, while the remaining infrared infinity part could be absorbed into the non-perturbative
universal inputs, such as the hadron wave functions.
The high-order QCD hard corrections  can then be extracted by making the difference of the full quark
amplitudes  with the convoluted amplitudes of effective wave functions and the hard amplitudes.

Over the past decade, the collinear factorization and the $\kt$ factorization for the exclusive processes
$\pi \gamma^{\star} \to \gamma(\pi)$, $B \to \gamma(\pi) l \overline{\nu}$ and $\rho \gamma^{\star} \to \pi$
have been demonstrated at the leading order (LO) and the next-to-leading
order (NLO) order \cite{PQCD-fact1,prd90-076001}.
And the NLO hard corrections to these processes have also been found in Refs.~\cite{PQCD-NLO-Li,PQCD-NLO-Cheng}.
Particularly, the NLO corrections to the time-like pion electromagnetic form factor are also calculated in
the PQCD approach \cite{PQCD-NLO-timelike},
which at the same time to improve the parameterization precision of the time-like pion form factor\cite{time-like pion ff},
also can help us to studies about the pion B meson three-body hadronic decays \cite{3b-others,3b-Wang,3b-AK}.

In this paper, we will consider the $\rho$ meson EM transition process, demonstrate the factorization for the
$\rho \gamma^* \to \rho$ transition at the NLO level, as an extension to the previous work \cite{prd90-076001}
where we verified explicitly that the factorization hypothesis is valid for the $\rho \gamma^* \to \pi$ process
at the NLO level. This work may help us to understand the exclusive processes involving at least a
vector hadron at the NLO level \cite{prd70-033001,jpg34-1845,prc65-045211,epjc67-253,prd92-094031}.

With the standard steps to factorize the full quark level amplitudes in the momentum, spin and color spaces,
we will find that the soft divergences will cancel in the quark level diagrams
and the remaining collinear divergences can be absorbed into the NLO wave functions. And we finally obtain
the infrared safe NLO correction amplitudes to the LO hard kernels as expected, and write down
these collinear part in the form of nonlocal two quarks hadron matrix element with the gauge factor Wilson line.

The paper is organized as following.
In section II we present the leading order dynamical analysis and the hard amplitudes of $\rho \to \rho$ transition process.
In section III we prove that the factorization hypothesis is valid for $\rho\gamma^* \to \rho$ transition
process at the next-to-leading order, and construct the NLO $\rho$ meson wave functions by means of the
nonlocal hadron matrix element with the gauge factor. A short summary will appear in the final section.

\section{Leading Order Hard Kernel}
\begin{figure}[tb]
\vspace{-1.5cm}
\begin{center}
\leftline{\epsfxsize=12cm\epsffile{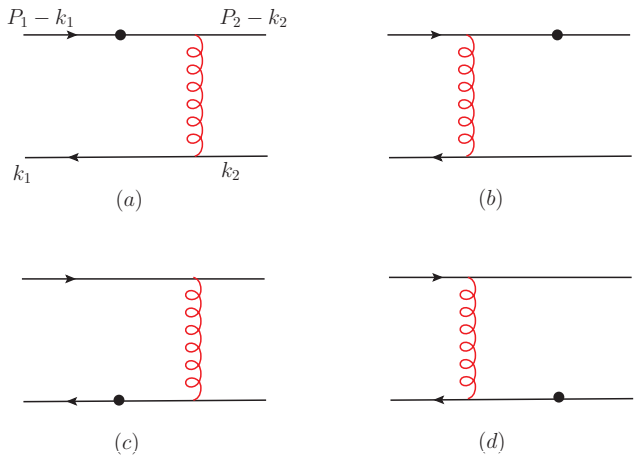}}
\end{center}
\vspace{-10cm}
\caption{The leading-order quark diagrams for the $\rho \gamma^{\star} \to \rho$ form factor,
with the virtual photon vertices, denoted by the symbol $\bullet$, setting at four different positions.}
\label{fig:fig1}
\end{figure}

The Leading-order quark diagrams for the $\rho \gamma^{\star} \to \rho$ transition are depicted in
Fig.~\ref{fig:fig1}, with the following kinematics defined in the light cone coordinate:
{\small
\beq
p_{1 \mu} &=& \frac{Q}{\sqrt{2}}(1,\gamma_{\rho}^2,\mathbf{0_T}), \quad
k_{1 \mu} = \frac{Q}{\sqrt{2}}(x_1,0,\mathbf{0_T});\non
\epsilon_{1 \mu}(L) &=& \frac{1}{\sqrt{2} \gamma_{\rho}}(1,-\gamma_{\rho}^2,\mathbf{0_{T}}),~~~~~
\epsilon_{1 \mu}(T)  =  (0,0,\mathbf{1_{T}});
\label{eq:kinematics-1}
\eeq
\beq
p_{2 \mu} &=& \frac{Q}{\sqrt{2}}(\gamma_{\rho}^2,1,\mathbf{0_T}),~~~~~
k_{2 \mu}  =  \frac{Q}{\sqrt{2}}(0,x_2,\mathbf{0_T});\non
\epsilon_{2 \mu}(L) &=& \frac{1}{\sqrt{2} \gamma_{\rho}}(-\gamma_{\rho}^2,1,\mathbf{0_{T}}),~~~~~
\epsilon_{2 \mu}(T)  =  (0,0,\mathbf{1_{T}}),
\label{eq:kinematics-2}
\eeq}
where the transverse momentum of the valance quarks in the initial and final mesons have been neglected.
The momentum fraction $x_i$ represent the strength for the corresponding valence quark to carry the longitudinal momentum,
and the polarization vectors $\epsilon_{i \mu}(L/T)$ is defined by the normalization conditions $\epsilon^2_{i}(L/T)=-1$, and
the parameter $\gamma_\rho=m_\rho/Q$.
With the above kinematics in hand,
the evolution behaviours of the $\rho$ meson radiation form factor on the lorentz invariant argument $Q^2$ can be obtained.
\beq
q^2 = (p_1-p_2)^2 \equiv -Q^2, \quad Q^2 > 0.
\label{eq:lorentz-invariant-QQ}
\eeq

The longitudinal and transverse $\rho$ meson wave functions are obtained from the light cone expansion around the
small interval $x^2 \sim 0$ and expressed in series of different twists \cite{WFs}.
By employing the kinematics as shown in Eqs.~(\ref{eq:kinematics-1},\ref{eq:kinematics-2}) and the
dimensionless unit vector $n=(1,0,\mathbf{0})$ and $v=(0,1,\mathbf{0})$ we can write the two-particle initial
and final $\rho$ meson wave functions as in Ref.~\cite{prd92-094031},
{\small
\beq
\Phi_{\rho}(p_1,\epsilon_{1T})&=&\frac{i}{\sqrt{2N_{c}}}
\left [M_{\rho} \esl_{1T} \phi^{v}_{\rho}(x_{1})+\esl_{1T} \psl_{1}\phi^{T}_{\rho}(x_{1})
+M_{\rho} i \epsilon_{\mu' \nu \rho \sigma} \gamma^{\mu'} \gamma_{5} \epsilon^{\nu}_{1T} n^{\rho} v^{\sigma} \phi^{a}_{\rho}(x_{1})\right ],\non
\Phi_{\rho}(p_1,\epsilon_{1L})&=&\frac{i}{\sqrt{2N_{c}}}\left [ M_{\rho} \esl_{1L} \phi_{\rho}(x_{1})
+\esl_{1L} \psl_{1}\phi^{t}_{\rho}(x_{1}) + M_{\rho} \phi^{s}_{\rho}(x_{1})\right ],
\label{eq:wfrho1}\\
\Phi_{\rho}(p_2,\epsilon_{2T})&=&\frac{i}{\sqrt{2N_{c}}}
\left [M_{\rho} \esl_{2T} \phi^{v}_{\rho}(x_{2})+\esl_{2T} \psl_{2}\phi^{T}_{\rho}(x_{2})
+M_{\rho} i \epsilon_{\mu' \nu \rho \sigma}  \gamma_{5}\gamma^{\mu'} \epsilon^{\nu}_{2T} v^{\rho} n^{\sigma} \phi^{a}_{\rho}(x_{2})\right ],\non
\Phi_{\rho}(p_2,\epsilon_{2L})&=&\frac{i}{\sqrt{2N_{c}}}\left [ M_{\rho} \esl_{2L} \phi_{\rho}(x_{2})
+\esl_{2L} \psl_{2}\phi^{t}_{\rho}(x_{2}) + M_{\rho} \phi^{s}_{\rho}(x_{2})\right ],
\label{eq:wfrho2}
\eeq}
where $N_c=3$ is the color number, the distribution amplitudes (DAs) $\phi_{\rho}$ and $\phi^{T}_{\rho}$ are the leading twist-2 (T2) terms,
while $\phi^{a/v}_{\rho}$ and $\phi^{t/s}_{\rho}$ are the sub-leading twist-3 (T3) terms.

It's very lucky that the four diagrams in Fig.~(\ref{fig:fig1}),with the different locations to radiate the virtual photon,
have the simple exchanging symmetry of the kinematics if we don't distinguish the flavor 'u' and 'd' in $\rho$ meson,
as claimed in \cite{PQCD-NLO-Li,PQCD-NLO-Cheng,prd92-094031}.
For the sake of simplicity, we can just make the calculation for one of these four diagrams in detail
and then extend the results to other three figures by simple exchanging symmetry.
We firstly concentrate on the study for Fig.~(\ref{fig:fig1})(a) in this paper.

In order to exactly show the factorization at NLO level and to obtain the
shapes of the NLO $\rho$ meson wave function term by term,
we here classify the LO transition amplitude $G^{(0)}_{a}$ according to the double expansions
to the different polarization components and the twists, for the initial and the final $\rho$ meson wave functions.
Then the LO amplitude with fixed polarization and twist is expressed as $G^{(0)}_{a,IJ,kl}$,
where the first subscript $a$ denotes the amplitude from Fig.~1(a), other subscripts $I,J=L,T(k,l=2,3)$
represent the polarization (twist) of the initial and the final meson  wave functions in the relevant calculation.

The independent LO amplitudes for $\rho \gamma^* \to \rho $ transition process can then be written in the following form,
{\small
\beq
G^{(0)}_{a,LL,22}(x_1;x_2) &=& -\frac{i e g^2_s C_F}{2} \mathrm{Tr}
\Bigl[  \frac{[\epsl_{1L} M_{\rho} \phi_{\rho}(x_1)]
\gamma^{\alpha}[\epsl_{2L} M_{\rho} \phi_{\rho}(x_2)]  \gamma_{\alpha}
(\psl_2 - \ksl_1) \gamma_{\mu}}{(p_2-k_1)^2 (k_1-k_2)^2} \Bigr],
\label{eq:loll22}
\eeq
\beq
G^{(0)}_{a,LL,33}(x_1;x_2) &=& -\frac{i e g^2_s C_F}{2} \mathrm{Tr}
\Bigl[  \frac{[ M_{\rho} \phi^{s}_{\rho}(x_1)]
\gamma^{\alpha}[ M_{\rho} \phi^{s}_{\rho}(x_2)]  \gamma_{\alpha}
(\psl_2 - \ksl_1) \gamma_{\mu}}{(p_2-k_1)^2 (k_1-k_2)^2} \Bigr],
\label{eq:loll33}
\eeq
\beq
G^{(0)}_{a,LT,23}(x_1;x_2) &=& -\frac{i e g^2_s C_F}{2} \mathrm{Tr} \Bigl[
\frac{[\epsl_{2T} M_{\rho} \phi^{v}_{\rho}(x_2)+ M_{\rho} i \epsilon_{\mu'\nu\rho\sigma}
\gamma_5 \gamma^{\mu'} \epsilon^{\nu}_{2T} v^{\rho} n^{\sigma} \phi^{a}_{\rho}(x_2)]}{(p_2-k_1)^2 (k_1-k_2)^2}\non
&&\cdot  \gamma_{\alpha} (\psl_2 - \ksl_1) \gamma_{\mu}[\epsl_{1L} M_{\rho} \phi_{\rho}(x_1)]
\gamma^{\alpha} \Bigr],
\label{eq:lolt23}
\eeq
\beq
G^{(0)}_{a,TL,23}(x_1;x_2) &=& -\frac{i e g^2_s C_F}{2} \mathrm{Tr}
\Bigl[ \frac{[M_{\rho} \phi^{s}_{\rho}(x_2)]  \gamma_{\alpha} (\psl_2 - \ksl_1) \gamma_{\mu}
[\epsl_{1T}\psl_1\phi^{T}_{\rho}(x_1)]\gamma^{\alpha}}{(p_2-k_1)^2 (k_1-k_2)^2} \Bigr],
\label{eq:lotl23}
\eeq
\beq
G^{(0)}_{a,TL,32}(x_1;x_2) &=& -\frac{i e g^2_s C_F}{2} \mathrm{Tr}
\Bigl[  \frac{[\epsl_{1T} M_{\rho} \phi^{v}_{\rho}(x_1)
+ M_{\rho} i \epsilon_{\mu'\nu\rho\sigma}\gamma^{\mu'} \gamma_5  \epsilon^{\nu}_{1T}
n^{\rho} v^{\sigma} \phi^{a}_{\rho}(x_1)]}{(p_2-k_1)^2 (k_1-k_2)^2}\non
&&\cdot \gamma^{\alpha}[\epsl_{2L} M_{\rho} \phi_{\rho}(x_2)]  \gamma_{\alpha} (\psl_2 - \ksl_1) \gamma_{\mu}\Bigr],
\label{eq:lotl32}
\eeq
\beq
G^{(0)}_{a,TT,33}(x_1;x_2) & =& -\frac{i e g^2_s C_F}{2} \mathrm{Tr}
\Bigl[ \frac{[\epsl_{1T} M_{\rho} \phi^{v}_{\rho}(x_1)+M_{\rho} i \epsilon_{\mu'\nu\rho\sigma}\gamma^{\mu'} \gamma_5
\epsilon^{\nu}_{1T} n^{\rho} v^{\sigma} \phi^{a}_{\rho}(x_1)]}{(p_2-k_1)^2 (k_1-k_2)^2}\non
&&\cdot \gamma^{\alpha} [\epsl_{2T} M_{\rho} \phi^{v}_{\rho}(x_2)+ M_{\rho} i \epsilon_{\mu'\nu\rho\sigma} \gamma_5
\gamma^{\mu'} \epsilon^{\nu}_{2T} v^{\rho} n^{\sigma} \phi^{a}_{\rho}(x_2)]
\gamma_{\alpha}(\psl_2 - \ksl_1) \gamma_{\mu} \Bigr],
\label{eq:lott33}
\eeq}
where $C_F=4/3$ is a color factor.
The standard analytical calculations for the LO amplitudes  in Eqs.~(\ref{eq:loll22}-\ref{eq:lott33}) show that:
\begin{enumerate}
\item [(1)]
For $G^{(0)}_{a,LL,22}$ and $G^{(0)}_{a,LL,33}$, where both the initial and final $\rho$ meson are longitudinal
polarized, only the wave functions with the same twist  power (T2 $\&$ T2, or T3 $\&$ T3)
for initial and
final states contribute to the radiation amplitude, since their is no helicity flip between outgoing and incoming
quarks for such combinations.
In the light cone coordinate, the matrix $\gamma_{\alpha}$ and $\gamma_{\mu}$ in Eq.~(\ref{eq:loll22})
are required to be $\gamma_{\bot}$ and $\gamma^{+}$ respectively.
While in Eq.~(\ref{eq:loll33}) with the T3 wave functions contribution,
the matrix $\gamma_{\alpha}$ is arbitrary and $\gamma_{\mu}$ is chosen to be $\gamma^{-}$
to collect the dominate contribution from $p^-_2$.

\item [(2)]
For the amplitudes with different polarizations of the initial and final $\rho$ meson,
the contributions only arise from different twist power of the initial and final wave functions( T2 $\& $ T3, or T3 $\&$ T2),
say the amplitude $G^{(0)}_{a,TL,23}$ and $G^{(0)}_{a,TL,32}$, because of the helicity flip.
For other possible such kind of combinations, the amplitude $G^{(0)}_{a,LT,23}$ can provide contribution, but
the amplitude $G^{(0)}_{a,LT,32}$ is forbidden due to the zero result of the contraction
$\gamma^{\alpha}\epsl_{2T}\psl_2\gamma_{\alpha}$.
It's clearly that the matrix $\gamma_{\alpha}$ is arbitrary in Eqs.~(\ref{eq:lolt23},\ref{eq:lotl23}),
while it should be $\gamma_{\bot}$ in Eq.~(\ref{eq:lotl32}).

\item [(3)]
For the amplitude with both transverse polarized initial and final $\rho$  meson,
say $G^{(0)}_{a,TT,22}$ and $G^{(0)}_{a,TT,33}$,  only $G^{(0)}_{a,TT,33}$ provide non-zero contribution, while
$G^{(0)}_{a,TT,22}$  does not because of the same reasons as mentioned above.
And the matrix $\gamma_{\alpha}$ in Eq.~(\ref{eq:lott33}) is arbitrary and $\gamma_{\mu}$ should to be
$\gamma^{-}$.
\end{enumerate}

\section{Factorization of $\rho \gamma^{\star} \to \rho$ at next-to-leading-order}

In this section we will firstly make the demonstration for the factorization of the exclusive electromagnetic
$\rho \gamma^{\star} \to \rho$ transition process at the NLO level  in the collinear and the soft approximations
and with the omission of  the transverse momentum.
At the end of this section, we will discuss how to factorize the infrared contribution from  the light cone
NLO meson wave functions safely in the $k_T$ factorization frame, where we will pick up the transverse momentum.

We again do not consider the infrared safe self-energy corrections to the far off-shell internal quark line,
as we have done for the $\rho \gamma^* \to \pi$ transition process \cite{prd90-076001}.
In general, the factorization of the long-distance and the short-distance physics in an exclusive QCD process
is not trivial.
To this end, we should divide the amplitudes simultaneously in the current and momentum spaces based
on their infrared properties,
and then sum up all the possible Feynman diagrams to collect the color factors and to maintain the gauge invariance.
In more detail, we can deal with the factorization process in three steps:
(a) Firstly, the Eikonal approximation is used to rearrange the singularity propagator into a more clear
formula without $\gamma$ matrix;
(b) Secondly, the Fierz identity in Eq.~(\ref{eq:fierz}) is used to separate the fermion currents,
the different terms on right-hand side (RHS) of Eq.~(\ref{eq:fierz}) can be treated as the contributions
with different twist, and then the infrared contribution obtained in step-I can be absorbed into
the newly defined wave functions;
and (c) we finally should consider all the possible diagrams for a gluon radiation,
which means that to collect the Feynman-diagrams with all possibilities for the end locations of the radiated gluon.
{\small
\beq
I_{ij}I_{lk}   &=& \frac{1}{4}I_{ik}I_{lj} + \frac{1}{4}(\gamma_{5})_{ik}(\gamma_{5})_{lj}
               + \frac{1}{4}(\gamma^{\alpha})_{ik}(\gamma^{\alpha})_{lj}\non
 &&              + \frac{1}{4}(\gamma_{5}\gamma^{\alpha})_{ik}(\gamma_{\alpha}\gamma_{5})_{lj}
               + \frac{1}{8}(\sigma^{\alpha\beta})_{ik}(\sigma_{\alpha\beta})_{lj},
\label{eq:fierz}
\eeq}

\subsection{Collinear factorization for the NLO(${\calo}(\alpha^2_s)$) corrections to Fig.1(a)}

\begin{figure}[tb]
  \centering
  \vspace{0cm}
  \begin{center}
  \leftline{\epsfxsize=14cm\epsffile{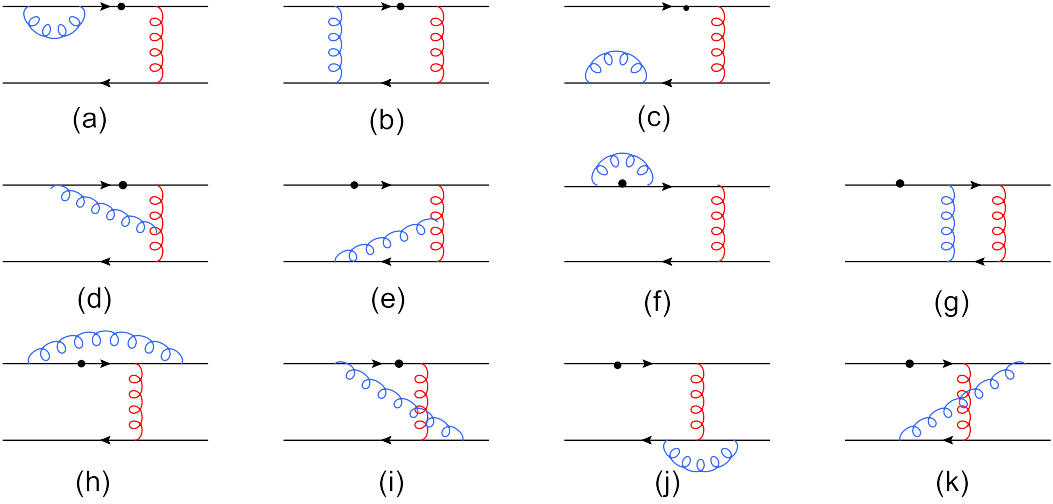}}
  \end{center}
  \vspace{-1cm}
\caption{The Feynman diagrams  which provide the NLO QCD corrections to Fig.~\ref{fig:fig1}(a)
with an additional gluon (blue curves) emitted from the quark (anti-quark) lines of the initial $\rho$ meson.}
\label{fig:fig2}
\end{figure}

We here will show the factorization of the NLO corrections for the Fig.~\ref{fig:fig1}(a) only,
The NLO corrections for other three relevant Feynman diagrams Figs.~\ref{fig:fig1}(b,c,d)
are similar in nature with those from Fig.~\ref{fig:fig1}(a) and can be obtained directly from the
result of Fig.~\ref{fig:fig1}(a) by simple kinematic replacements.
The NLO corrections to Fig.~\ref{fig:fig1}(a), in principle, can be divided into two parts:
\ben
\item[(1)] The first set contains the diagrams where the gluon lines emitted from the valence quark lines of the initial $\rho$ meson
and attached to any other possible places, as depicted in Fig.~\ref{fig:fig2}.

\item[(2)] The second set has the diagrams where the gluon lines emitted from the valence quark lines
of the final $\rho$ meson, very similar with those as illustrated in Fig.~\ref{fig:fig2}.
\een

We therefore will firstly deal with Fig.~\ref{fig:fig2}, where the emitted gluon may be parallel to
the initial $\rho$ meson momentum $p_1$.
Even more, in a previous paper \cite{prd90-076001}
the transversely polarized NLO initial $\rho$ meson wave functions $\Phi^T_{\rho}(x_1)$ (T2) and
$\Phi^{v,a}_{\rho}(x_1)$ (T3) have been defined in a form of non-local matrix elements and studied systematically.
So here we can consider only for the longitudinally polarized initial $\rho$ meson wave functions
$\Phi_{\rho}(x_1)$ (T2) and $\Phi^s_{\rho}(x_1)$ (T3) which entered into Fig.~\ref{fig:fig2} through
Eqs.~(\ref{eq:loll22},\ref{eq:loll33}).
Since the $\rho$ meson wave function $\Phi^t_{\rho}(x_1)$ (T3) does not contribute through Fig.~\ref{fig:fig1} at the
leading order, we can't get much NLO knowledge about this wave function $\Phi^t_{\rho}(x_1)$.

In general, there exist two kinds of standard infrared divergence
\footnote{A third kind infrared divergence arose from the Glauber region where $l \sim (\lambda^2/Q, \lambda^2/Q, \lambda)$
\cite{prd90-074018},
and this Glauber divergence appeared in the NLO spectator amplitudes of B meson two-body nonleptonic decays,
can help us to resolve the long-standing $\pi\pi$ puzzle while still survive the constraints in the standard model
\cite{prd91-114019}.}
for the Feynman diagrams in Fig.~\ref{fig:fig2}.
When the emitted gluon is exchanged between two on-shell external lines,
the soft divergence will appear when the momentum $l = (l^+, l^-, l_{\bot})\sim (\lambda, \lambda, \lambda)$ (here $\lambda \sim \Lambda_{QCD}$).
And when the collinear gluon $l \sim (Q, \lambda^2/Q, \lambda^2)$ is emitted from a massless external quark line,
this gluon will generate the collinear divergence \cite{jhep0802-002,npb844-199}.
In the $\rho$ meson electromagnetic transition process,
both the external and internal quark lines are light quarks and can be considered as massless,
so the collinear divergence can't be regularized by the quark mass and have to be absorbed into the meson wave functions,
which is one of the main purposes of this paper as mentioned previously.

It's straightforward to write down the NLO amplitudes of Figs.~\ref{fig:fig2}(a,b,c)
because the end location of the radiated gluon lines don't attach to the internal lines.
Then their radiations don't pollute the LO hard kernel even for the format of the corresponding amplitudes,
so that such  amplitudes can be factorized by a simple insertion of the Fierz identity
as shown in Eq.~(\ref{eq:fierz}) in the external quark lines,
{\small
\beq
G^{(1)}_{2a,LL,22} &=& -\frac{1}{2} \frac{e g^4_s C^2_F}{2} \mathrm{Tr}\Bigl[
    \frac{[\epsl_{1L} M_{\rho} \phi_{\rho}(x_1)]\gamma^{\alpha} [\epsl_{2L} M_{\rho} \phi_{\rho}(x_2)]\gamma_{\alpha}
    (\psl_2- \ksl_1) \gamma_{\mu} }{(p_2-k_1)^2 (k_1-k_2)^2 (p_1-k_1)^2 (p_1-k_1+l)^2 l^2} \non
&& \cdot (\psl_1-\ksl_1) \gamma^{\rho'}(\psl_1-\ksl_1+\lsl)\gamma_{\rho'}\Bigr]\non
&=& \frac{1}{2}\Phi^{(1)}_{\rho, a}(x_1,\xi_1) \otimes G^{(0)}_{a,LL,22}(x_1; x_2),
\label{eq:nloall22}
\eeq
\beq
G^{(1)}_{2b,LL,22} &=& \frac{ e g^4_s C^2_F}{2} \mathrm{Tr} \Bigl[
    \frac{[\epsl_{1L} M_{\rho} \phi_{\rho}(x_1)] \gamma^{\rho'}(\ksl_1-\lsl)\gamma_{\alpha} [\epsl_{2L} M_{\rho} \phi_{\rho}(x_2)]
    \gamma_{\alpha}} {(p_2-k_1+l)^2 (k_1-k_2-l)^2 (p_1-k_1+l)^2 (k_1-l)^2 l^2} \non
&&\cdot (\psl_2-\ksl_1+\lsl) \gamma_{\mu} (\psl_1-\ksl_1+\lsl) \gamma_{\rho'}\Bigr]\non
&=&\Phi^{(1)}_{\rho, b}(x_1,\xi_1) \otimes G^{(0)}_{a,LL,22}(\xi_1; x_2),
\label{eq:nlobll22}\\
G^{(1)}_{2c,LL,22} &=&- \frac{1}{2} \frac{e g^4_s C^2_F}{2} \mathrm{Tr} \Bigl[
    \frac{[\epsl_{1L} M_{\rho} \phi_{\rho}(x_1)] \gamma^{\rho'} (\ksl_1-\lsl)\gamma_{\rho'}\ksl_1\gamma^{\alpha}[\epsl_{2L} M_{\rho} \phi_{\rho}(x_2)] \gamma_{\alpha} (\psl_2-\ksl_1)\gamma_{\mu} }
    {(p_2-k_1)^2 (k_1-k_2)^2 (k_1-l)^2 (k_1)^2 l^2} \Bigr]\non
&=& \frac{1}{2}\Phi^{(1)}_{\rho, c}(x_1,\xi_1) \otimes G^{(0)}_{a,LL,22}(x_1; x_2),
\label{eq:nlocll22}\\
G^{(1)}_{2a,LL,33} &=& -\frac{1}{2} \frac{e g^4_s C^2_F}{2} \mathrm{Tr} \Bigl[
    \frac{[M_{\rho}\phi^s_{\rho}(x_1)]\gamma^{\alpha}[M_{\rho}\phi^s_{\rho}(x_2)]\gamma_{\alpha}(\psl_2 - \ksl_1)\gamma_{\mu}
    (\psl_1 - \ksl_1)\gamma^{\rho'} (\psl_1 - \ksl_1 + \lsl) \gamma_{\rho'}}
    {(p_2-k_1)^2 (k_1-k_2)^2 (p_1-k_1)^2 (p_1-k_1+l)^2 l^2}  \Bigr] \non
&=& \frac{1}{2} \Phi^{(1),s}_{\rho, a}(x_1,\xi_1) \otimes G^{(0)}_{a,LL,33}(x_1; x_2),
\label{eq:nloall33}\\
G^{(1)}_{2b,LL,33} &=& \frac{ e g^4_s C^2_F}{2} \mathrm{Tr}\Bigl[
    \frac{[M_{\rho}\phi^s_{\rho}(x_1)]\gamma^{\rho'} (\ksl_1 - \lsl) \gamma^{\alpha} [ M_{\rho} \phi^s_{\rho}(x_2)] \gamma_{\alpha}
    (\psl_2 - \ksl_1 + \lsl)\gamma_{\mu}(\psl_1 - \ksl_1 + \lsl) \gamma^{\rho'}}
    {(p_2-k_1+l)^2 (k_1-k_2-l)^2 (p_1-k_1+l)^2 (k_1-l)^2 l^2} \Bigr]\non
&=& \Phi^{(1),s}_{\rho, b}(x_1, \xi_1) \otimes G^{(0)}_{a,LL,33}(\xi_1; x_2),
\label{eq:nlobll33}\\
G^{(1)}_{2c,LL,33} &=&- \frac{1}{2} \frac{e g^4_s C^2_F}{2} \mathrm{Tr}\Bigl[
    \frac{[M_{\rho}\phi^s_{\rho}(x_1)] \gamma^{\rho'} (\ksl_1-\lsl)\gamma_{\rho'}\ksl_1\gamma^{\alpha}[M_{\rho}\phi^s_{\rho}(x_2)]
    \gamma_{\alpha} (\psl_2-\ksl_1)\gamma_{\mu} }
    {(p_2-k_1)^2 (k_1-k_2)^2 (k_1-l)^2 (k_1)^2 l^2} \Bigr]\non
&=& \frac{1}{2}\Phi^{(1)}_{\rho, c}(x_1,\xi_1)\otimes G^{(0)}_{a,LL,33}(x_1; x_2)  ,
\label{eq:nlocll33}
\eeq }
where the NLO wave functions $\Phi^{(1)}_{\rho, i}$ and $\Phi^{(1),s}_{\rho, i}$ with $i=(a,b,c)$ will absorb
all the infrared singularities from those reducible sub-diagrams Figs.~\ref{fig:fig2}(a,b,c), and can be written in
the form of
{\small
\beq
\Phi^{(1)}_{\rho, a}(x_1,\xi_1)&=& \frac{- i g^2_s C_F}{4} \mathrm{Tr} \Bigl[
\frac{\gamma^-_{\rho} \gamma^+_{\rho} (\psl_1-\ksl_1) \gamma^{\rho'} (\psl_1-\ksl_1+\lsl)
\gamma_{\rho'}}{(p_1-k_1)^2 (p_1-k_1+l)^2 l^2} \Bigr],  \non
\Phi^{(1)}_{\rho, b}(x_1,\xi_1) &=& \frac{i g^2_s C_F}{4} \mathrm{Tr} \Bigl[
\frac{\gamma^-_{\rho}\gamma^{\rho'}(\ksl_1-\lsl) \gamma^+_{\rho}  (\psl_1-\ksl_1+\lsl)\gamma_{\rho'}}
 {(p_1-k_1+l)^2 (k_1-l)^2 l^2} \Bigr],  \non
\Phi^{(1)}_{\rho, c}(x_1,\xi_1) &=& \frac{- i g^2_s C_F}{4} \mathrm{Tr} \Bigl[
\frac{ \gamma^-_{\rho} \gamma^{\rho'} (\ksl_1-\lsl)\gamma_{\rho'}\ksl_1 \gamma^+_{\rho}}
{(k_1-l)^2 (k_1)^2 l^2}\Bigr],
\label{eq:nlorho-abc}\\
\Phi^{(1),s}_{\rho, a}(x_1,\xi_1) &=& \frac{- i g^2_s C_F}{4} \mathrm{Tr}\Bigl[
\frac{(\psl_1-\ksl_1) \gamma^{\rho'} (\psl_1-\ksl_1+\lsl)
\gamma_{\rho'}}{(p_1-k_1)^2 (p_1-k_1+l)^2 l^2} \Bigr], \non
\Phi^{(1),s}_{\rho, b}(x_1,\xi_1) &=& \frac{i g^2_s C_F}{4} \mathrm{Tr} \Bigl[
\frac{\gamma^{\rho'}(\ksl_1-\lsl) (\psl_1-\ksl_1+\lsl)\gamma_{\rho'}}
 {(p_1-k_1+l)^2 (k_1-l)^2 l^2} \Bigr], \non
\Phi^{(1),s}_{\rho, c}(x_1,\xi_1)&=& \frac{- i g^2_s C_F}{4} \mathrm{Tr} \Bigl[
\frac{ \gamma^{\rho'} (\ksl_1-\lsl)\gamma_{\rho'}\ksl_1}
{(k_1-l)^2 (k_1)^2 l^2} \Bigr].
\label{eq:nlorhos-abc}
\eeq}
The QCD dynamics requires that the soft gluon could not resolve the color structure of the initial $\rho$ meson,
so it's reasonable to find that the soft divergences in these reducible amplitudes
$G^{(1)}_{2a,LL,kl}, G^{(1)}_{2b,LL,kl}, G^{(1)}_{2c,LL,kl}$ with $k=(l,2,3)$,
corresponding to the Figs.~(\ref{fig:fig2})(a,b,c), will  be canceled from each other.
And the remaining collinear divergences can then be absorbed into the NLO wave functions $\Phi^{(1)}_{\rho, j}$
and $\Phi^{(1),s}_{\rho, j}$ with $j=(a,b,c)$.

We then consider the NLO correction to the LO amplitude $G^{(0)}_{a,LL,22}$ in Eq.~(\ref{eq:loll22}).
The irreducible sub-diagrams Figs.~\ref{fig:fig2}(d-g) will generate collinear divergences  only,
Since the soft contributions, generated when the end locations of the radiated gluons are connected
to one of the internal propagators, are highly suppressing by the factor $1/Q^2$.
What's more, the infrared divergence in Figs.~\ref{fig:fig2}(f,g) are suppressed also by the kinematics.
We therefore can set their amplitudes $G^{(1)}_{2f,LL,22}$ and $G^{(1)}_{2g,LL,22}$ as zero safely in the
factorization theorem.
{\small
\beq
G^{(1)}_{2d,LL,22} &=& - e g^4_s  \mathrm{Tr}
\Bigl[ \frac{[\epsl_{1L} M_{\rho} \phi_{\rho}(x_1)] \gamma^{\alpha}[\epsl_{2L} M_{\rho} \phi_{\rho}(x_2)]
    \gamma^{\beta}(\psl_2-\ksl_1+\lsl)\gamma_{\mu}(\psl_1-\ksl_1+\lsl) \gamma^{\gamma} F_{\alpha \beta \gamma}}
    {(p_2-k_1+l)^2 (k_1-k_2)^2 (k_1-k_2-l)^2 (p_1-k_1+l)^2 l^2} \Bigr] \non
  &\sim & \frac{9}{8} \Phi^{(1)}_{\rho, d}(x_1,\xi_1) \otimes \left [G^{(0)}_{a,LL,22}(x_1, \xi_1; x_2) - G^{(0)}_{a,LL,22}(\xi_1; x_2)\right],
\label{eq:nlodll22} \\
G^{(1)}_{2e,LL,22} &=&  e g^4_s \mathrm{Tr}
\Bigl[ \frac{[\epsl_{1L} M_{\rho} \phi_{\rho}(x_1)] \gamma^{\gamma}(\ksl_1-\lsl)\gamma^{\alpha}
    [\epsl_{2L} M_{\rho} \phi_{\rho}(x_2)] \gamma^{\beta} (\psl_2-\ksl_1) \gamma_{\mu} F_{\alpha \beta \gamma}}
    {(p_2-k_1)^2 (k_1-k_2)^2 (k_1-k_2-l)^2 (k_1-l)^2 l^2}\Bigr] \non
  &\sim & \frac{9}{8} \Phi^{(1)}_{\rho, e}(x_1,\xi_1) \otimes \left [G^{(0)}_{a,LL,22}(x_1; x_2) - G^{'(0)}_{a,LL,22}(x_1, \xi_1; x_2) \right],
\label{eq:nloell22}\\
G^{(1)}_{2f,LL,22} &=& \frac{-e g^4_s C^2_F }{2} \mathrm{Tr}
\Bigl[ \frac{[\epsl_{1L} M_{\rho} \phi_{\rho}(x_1)] \gamma^{\alpha}[\epsl_{2L} M_{\rho} \phi_{\rho}(x_2)] \gamma_{\alpha}(\psl_2-\ksl_1) \gamma^{\rho'}(\psl_2-\ksl_1+\lsl)}
    {(p_2-k_1)^2 (k_1-k_2)^2 (p_2-k_1+l)^2 l^2 (p_1-k_1+l)^2} \non
&& \cdot \gamma_{\mu}(\psl_1-\ksl_1+\lsl) \gamma_{\rho'} \Bigr]\non
&\sim & 0,
\label{eq:nlofll22}\\
G^{(1)}_{2g,LL,22} &= &\frac{ e g^4_s C^2_F}{2} \mathrm{Tr} \Bigl[
    \frac{[\epsl_{1L} M_{\rho} \phi_{\rho}(x_1)] \gamma^{\rho'}(\ksl_1-\lsl)\gamma^{\alpha}[\epsl_{2L} M_{\rho} \phi_{\rho}(x_2)]\gamma_{\alpha}(\psl_2-\ksl_1+\lsl) \gamma_{\rho'}}
    {(p_2-k_1)^2 (k_1-k_2-l)^2 (k_1-l)^2 l^2 (p_2-k_1+l)^2} \non
&& \cdot  (\psl_2-\ksl_1)\gamma_{\mu} \Bigr] \non
 & \sim & 0,
\label{eq:nlogll22}
\eeq}
The new LO hard amplitudes in Eqs.~(\ref{eq:nlodll22},\ref{eq:nloell22}) with different variables $x_1, \xi_1$, and $x_2$
are collected as following,
{\small
\beq
G^{(0)}_{a,LL,22}(\xi_1;x_2)&=&\frac{-i e g^2_s C_F}{2} \mathrm{Tr}
\Bigl[ \frac{[\epsl_{1L} M_{\rho} \phi_{\rho}(x_1)] \gamma^{\alpha} [\epsl_{2L} M_{\rho} \phi_{\rho}(x_2)]\gamma_{\alpha}(\psl_2 - \ksl_1+\lsl ) \gamma_{\mu}}
    {(p_2-k_1+l)^2 (k_1-k_2-l)^2} \Bigr],
\label{eq:loall221} \\
G^{(0)}_{a,LL,22}(x_1, \xi_1; x_2)&=&\frac{-i e g^2_s C_F}{2} \mathrm{Tr}
\Bigl[  \frac{[\epsl_{1L} M_{\rho} \phi_{\rho}(x_1)] \gamma^{\alpha} [\epsl_{2L} M_{\rho} \phi_{\rho}(x_2)]\gamma_{\alpha}(\psl_2 - \ksl_1+\lsl ) \gamma_{\mu}}
    {(p_2-k_1+l)^2 (k_1-k_2)^2} \Bigr],
\label{eq:loall222} \\
G^{'(0)}_{a,LL,22}(x_1, \xi_1; x_2)&=&\frac{-i e g^2_s C_F}{2} \mathrm{Tr}
\Bigl[ \frac{[\epsl_{1L} M_{\rho} \phi_{\rho}(x_1)] \gamma^{\alpha} [\epsl_{2L} M_{\rho} \phi_{\rho}(x_2)]\gamma_{\alpha}(\psl_2 - \ksl_1 ) \gamma_{\mu}}
    {(p_2-k_1)^2 (k_1-k_2-l)^2} \Bigr].
\label{eq:loall223}
\eeq}
And it's easy to find the relation $G^{(0)}_{a,LL,22}(x_1; \xi_1, x_2)=G^{'(0)}_{a,LL,22}(x_1; \xi_1, x_2)$
in the collinear region $l \parallel p_1$.
For the tensor $F_{\alpha \beta \gamma} = g_{\alpha \beta} (2k_2-2k_1+l)_{\gamma} + g_{\beta \gamma} (k_1-k_2-2l)_{\alpha}
+g_{\gamma \alpha} (k_1-k_2+l)_{\beta}$ in Eq.~(\ref{eq:nlodll22}),
only the terms proportional to $g_{\alpha \beta}$ provide the corrections to the LO hard kernel $G^{(0)}_{a,LL,22}$.
After applying the eikonal approximation to divide the collinear divergence out of the several LO hard kernels $G^{(0)}$,
the NLO twist-2 longitudinal $\rho$ meson wave function $\Phi^{(1)}_{\rho, d}$, with the gluon radiated from the left-up quark line,
can be written as
{\small
\beq
\Phi^{(1)}_{\rho, d}(x_1,\xi_1) = \frac{- i g^2_s C_F}{4} \mathrm{Tr}
\Bigl[\frac{[\gamma^-_{\rho} \gamma^+_{\rho}](\psl_1-\ksl_1+\lsl) \gamma^{\rho'}  v_{\rho'}}
    {(p_1-k_1+l)^2 l^2 (v\cdot l)} \Bigr].
\label{eq:nlorho-d}
\eeq}
Similarly, for the tensor $F_{\alpha \beta \gamma} = g_{\alpha \beta} (2k_2-2k_1+l)_{\gamma} + g_{\beta \gamma} (k_1-k_2+l)_{\alpha}
+g_{\gamma \alpha} (k_1-k_2-2l)_{\beta}$ in Eq.~(\ref{eq:nloell22}) , only the first term contributes to the LO hard kernel
$G^{(0)}_{a,LL,22}$. And we can also write the NLO twist-2 longitudinal $\rho$ meson wave function $\Phi^{(1)}_{\rho, e}$,
in which the additional gluon is emitted from the left-down anti-parton line, in the following form,
{\small
\beq
\Phi^{(1)}_{\rho, e}(x_1,\xi_1) = \frac{i g^2_s C_F}{4} \mathrm{Tr}
\Bigl[    \frac{[\gamma^-_{\rho}] \gamma^{\rho'} (\ksl_1-\lsl) [\gamma^+_{\rho}] v_{\rho'}}
    {(k_1-l)^2 l^2 (v\cdot l)}\Bigr].
\label{eq:nlorho-e}
\eeq}

The remaining sub-diagrams in Figs.~\ref{fig:fig2}(h-k) may include the collinear divergence
as well as the soft divergence. The corresponding NLO amplitudes can be  written in the following standard forms,
{\small
\beq
G^{(1)}_{2h,LL,22} &=& \frac{ e g^4_s }{9} \mathrm{Tr}
\Bigl[ \frac{[\epsl_{1L} M_{\rho} \phi_{\rho}(x_1)]\gamma^{\alpha}[\epsl_{2L} M_{\rho} \phi_{\rho}(x_2)]
    \gamma^{\rho'}(\psl_2-\ksl_2+\lsl)\gamma_{\alpha} }{(p_2-k_1+l)^2 (k_1-k_2)^2 (p_2-k_2+l)^2 l^2 (p_1-k_1+l)^2 } \non
&& \cdot (\psl_2-\ksl_1+\lsl) \gamma_{\mu} (\psl_1-\ksl_1+\lsl)\gamma_{\rho'}\Bigr] \non
 & \sim &-\frac{1}{8}\Phi^{(1)}_{\rho, d}(x_1,\xi_1) \otimes G^{(0)}_{a,LL,22}(x_1, \xi_1; x_2) ,
\label{eq:nlohll22}\\
G^{(1)}_{2i,LL,22} &=& \frac{- e g^4_s }{9} \mathrm{Tr} \Bigl[
\frac{[\epsl_{1L} M_{\rho} \phi_{\rho}(x_1)]\gamma^{\alpha}(\ksl_2+\lsl)\gamma^{\rho'}[\epsl_{2L}
M_{\rho} \phi_{\rho}(x_2)]\gamma_{\alpha}}{(p_2-k_1+l)^2(k_1-k_2-l)^2 (p_1-k_1+l)^2 l^2 (k_2+l)^2} \non
&& \cdot (\psl_2-\ksl_1+\lsl) \gamma_{\mu} (\psl_1-\ksl_1+\lsl)\gamma_{\rho'}\Bigr] \non
 & \sim & \frac{1}{8}\Phi^{(1)}_{\rho, d}(x_1,\xi_1) \otimes G^{(0)}_{a,LL,22}(\xi_1; x_2),
\label{eq:nloill22}\\
G^{(1)}_{2j,LL,22} &=& \frac{e g^4_s }{9} \mathrm{Tr} \Bigl[
    \frac{[\epsl_{1L} M_{\rho} \phi_{\rho}(x_1)]\gamma_{\rho'} (\ksl_1-\lsl) \gamma^{\alpha}(\ksl_2-\lsl)\gamma^{\rho'}[\epsl_{2L} M_{\rho} \phi_{\rho}(x_2)]}{(p_2-k_1)^2 (k_1-k_2)^2 (k_2-l)^2 l^2 (k_1-l)^2}\non
&& \cdot \gamma_{\alpha} (\psl_2-\ksl_1) \gamma_{\mu} \Bigr] \non
&  \sim & -\frac{1}{8}\Phi^{(1)}_{\rho, e}(x_1,\xi_1) \otimes G^{(0)}_{a,LL,22}( x_1; x_2) ,
\label{eq:nlojll22}\\
G^{(1)}_{2k,LL,22} &=& \frac{- e g^4_s }{9} \mathrm{Tr} \Bigl[
    \frac{[\epsl_{1L} M_{\rho} \phi_{\rho}(x_1)]\gamma^{\rho'}(\ksl_1-\lsl)\gamma^{\alpha}[\epsl_{2L} M_{\rho} \phi_{\rho}(x_2)]\gamma_{\rho'}
}{(p_2-k_1)^2 (k_1-k_2-l)^2 (k_1-l)^2 l^2 (p_2-k_2-l)^2} \non
&& \cdot (\psl_2-\ksl_2-\lsl) \gamma_{\alpha} (\psl_2-\ksl_1)\gamma_{\mu}\Bigr] \non
 & \sim &\frac{1}{8}\Phi^{(1)}_{\rho, e}(x_1,\xi_1) \otimes G^{'(0)}_{a,LL,22}( x_1,\xi_1; x_2).
\label{eq:nlokll22}
\eeq }

To investigate the collinear factorization in the NLO level while keeping the gauge invariance,
we should sum up all the irreducible amplitudes with the same gluon radiation starting-point together.
The summed amplitudes for the Feynman diagrams with the gluon radiated from the left-up quark line can be written as
{\small
\beq
G^{(1)}_{2up,LL,22} (x_1; x_2)&=& \sum_{m=d,f,h,i}G^{(1)}_{2m,LL,22}(x_1; x_2) \non
  &=& \Phi^{(1)}_{\rho,d}(x_1,\xi_1) \otimes \left[G^{(0)}_{a,LL,22}(x_1,\xi_1; x_2) -G^{(0)}_{a,LL,22}(\xi_1; x_2)\right].
  \label{eq:nloupll22}
\eeq}
While the summed amplitudes for the gluon radiated from the left-down anti-quark line is written as
{\small
\beq
G^{(1)}_{2down,LL,22}(x_1; x_2) &=& \sum_{n=e,g,j,k}G^{(1)}_{2n,LL,22}(x_1; x_2) \non
  &=& \Phi^{(1)}_{\rho,e}(x_1,\xi_1) \otimes \left[G^{(0)}_{a,LL,22}(x_1; x_2)- G^{'(0)}_{a,LL,22}(x_1,\xi_1; x_2)\right].
\label{eq:nlodownll22}
\eeq}
There exist no soft divergences in these summed amplitudes.
With a simple deformation of the amplitudes in the soft kinematic region,
the soft singularities in Eqs.~(\ref{eq:nlohll22},\ref{eq:nloill22}) and Eqs.~(\ref{eq:nlojll22},\ref{eq:nlokll22}) cancel
by themselves.
The remaining collinear divergences, fortunately, can be absorbed into the NLO T2 $\rho$ meson
Wave function $\Phi^{(1)}_{\rho}$ by re-definition  of the nonlocal hadronic matrix element with the spin structure $\gamma^-/2$:
{\small
\beq
\Phi^{(1)}_{\rho}&=&\frac{1}{2N_c P^+_1} \int \frac{dy^-}{2\pi} e^{-i x p^+_1 y^-}
    <0| \overline{q}(y^-) \frac{\gamma^-}{2} (-i g_s)  \int^{y^-}_{0} dzv
    A(zv) q(0)| \rho(p_1)>.
\label{eq:nlo-rho}
\eeq}

  \begin{figure}[tb]
  \centering
  \vspace{0cm}
  \begin{center}
  \leftline{\epsfxsize=16cm\epsffile{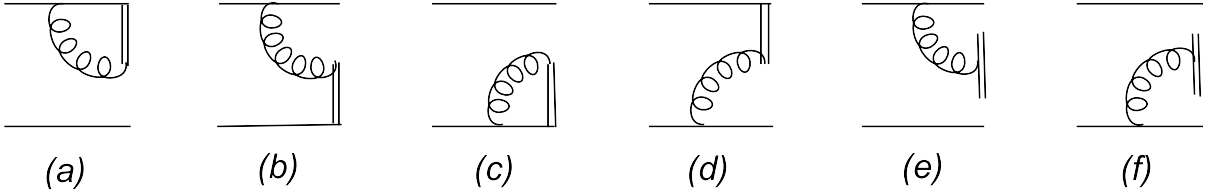}}
  \end{center}
  \vspace{-1cm}
  \caption{${\calo}(\alpha_s)$ effective diagrams for the initial longitudinal $\rho$ meson wave function.
  The vertical double line denotes the Wilson line along the light cone, whose Feynman rule is
  $v_\rho/(v \cdot l)$ as described in Eqs.~(\ref{eq:nlo-rho},\ref{eq:nlo-rhos},\ref{eq:nlo-rhoT},\ref{eq:nlo-rhov},\ref{eq:nlo-rhoa}).}
  \label{fig:fig3}
  \end{figure}

The relevant effective diagrams for the NLO initial $\rho$ meson wave function are shown in Fig.~\ref{fig:fig3},
in which all the collinear divergences from the rho meson radiation in the irreducible NLO quark diagrams
Fig.~\ref{fig:fig2}(d-k) are collected.
Figs.~\ref{fig:fig3}(a,c) describe the radiative corrections with the gluon momentum not flow into the LO hard kernel.
Figs.~\ref{fig:fig3}(b,d) ( Figs.~\ref{fig:fig3}(e,f) )  lead to the radiative corrections with the gluon momentum
flow into the LO hard kernel completely ( partially).
For the NLO vector wave functional $\Phi^{(1)}_{\rho}$ discussed in this paper,
only the sub-diagrams Figs.~\ref{fig:fig3}(b,c,e,f) give contributions.
This is because, as described in Eqs.~(\ref{eq:nloupll22},\ref{eq:nlodownll22}),
the gluon momentum started at the up-line quark flow into the LO hard kernel ( no matter completely or partially),
while the gluon momentum started at the down-line anti-quark can flow into the LO hard kernel partially or don't flow into
the LO hard kernel.
Then the collinear factorization hypothesis is valid for the NLO corrections to the LO amplitude $G^{(0)}_{a,LL,22}$
in Eq.~(\ref{eq:loll22}) with the longitudinally  polarized and the leading twist initial and final states
wave functions $\Phi_{\rho}$.

For the NLO corrections to the LO amplitude $G^{(0)}_{a,LL,33}$ in Eq.~(\ref{eq:loll33}),
the infrared divergences in the NLO amplitudes corresponding to Figs.~\ref{fig:fig2}(d,f,h,i) are all forbidden by kinematics.
Then only the sub-diagrams Figs.~\ref{fig:fig2}(e,g,j,k) provide the infrared corrections to the LO amplitude
$G^{(0)}_{a,LL,33}$. The relevant NLO amplitudes are of the following form,
{\small
\beq
G^{(1)}_{2d,LL,33} &=& e g^4_s \mathrm{Tr}\Bigl[
    \frac{[M_{\rho}\phi^s_{\rho}(x_1)] \gamma^{\alpha} [M_{\rho} \phi^s_{\rho}(x_2)] \gamma_{\beta} (\psl_2 - \ksl_1+ \lsl) \gamma^{\mu}
    (\psl_1 - \ksl_1 + \lsl)\gamma^{\gamma} F_{\alpha \beta \gamma}}{(p_2-k_1+l)^2
    (k_1-k_2)^2 (p_1-k_1+l)^2 (k_1-k_2-l)^2 l^2}\Bigr]\non
& \sim & 0,
\label{eq:nlodll33}\\
G^{(1)}_{2e,LL,33} &= & - e g^4_s \mathrm{Tr}\Bigl[
    \frac{[M_{\rho}\phi^s_{\rho}(x_1)] \gamma^{\gamma} (\ksl_1 - \lsl) \gamma^{\alpha} [M_{\rho} \phi^s_{\rho}(x_2)] \gamma_{\beta} (\psl_2 - \ksl_1)
    \gamma^{\mu} F_{\alpha \beta \gamma} }{(p_2-k_1)^2 (k_1-k_2)^2 (k_1-l)^2 (k_1-k_2-l)^2 l^2} \Bigr]\non
& \sim & \frac{9}{16} \Phi^{(1),s}_{\rho, e}(x_1,\xi_1) \otimes [G^{(0)}_{a,LL,33}(x_1; x_2) - G^{(0)}_{a,LL,33}(x_1,\xi_1;x_2)],
\label{eq:nloell33}\\
G^{(1)}_{2f,LL,33}& = & \frac{e g^4_s}{9} \mathrm{Tr}\Bigl[
\frac{[M_{\rho}\phi^s_{\rho}(x_1)]\gamma^{\alpha} [M_{\rho} \phi^s_{\rho}(x_2)]
\gamma_{\alpha} (\psl_2 - \ksl_1) \gamma^{\rho'}(\psl_2-\ksl_1+\lsl) \gamma_{\mu}
(\psl_1-\ksl_1+\lsl) \gamma_{\rho'}}{(p_2-k_1)^2 (k_1-k_2)^2 (p_1-k_1+l)^2 (p_2-k_1+l)^2 l^2}\Bigr] \non
& \sim & 0,
\label{eq:nlofll33}
\eeq
\beq
G^{(1)}_{2g,LL,33} &= & \frac{- e g^4_s}{9} \mathrm{Tr}\Bigl[
\frac{[M_{\rho}\phi^s_{\rho}(x_1)]\gamma_{\rho'} (\ksl_1 - \lsl) \gamma^{\alpha} [M_{\rho} \phi^s_{\rho}(x_2)] \gamma_{\alpha}
(\psl_2 - \ksl_1+\lsl)\gamma^{\rho'}(\psl_2-\ksl_1) \gamma_{\mu}}{(p_2-k_1)^2
(k_1-k_2-l)^2 (p_2-k_1+l)^2 (k_1-l)^2 l^2}  \Bigr]\non
&\sim &  \Phi^{(1),s}_{\rho, e}(x_1,\xi_1) \otimes [G^{'(0)}_{a,LL,33}(x_1,\xi_1; x_2) - G^{(0)}_{a,LL,33}(\xi_1;x_2)] .
\label{eq:nlogll33}\\
G^{(1)}_{2h,LL,33} &= & \frac{e g^4_s }{9} \mathrm{Tr}\Bigl[
    \frac{[M_{\rho}\phi^s_{\rho}(x_1)]\gamma^{\alpha} [M_{\rho} \phi^s_{\rho}(x_2)] \gamma^{\rho'}
    (\psl_2-\ksl_2+\lsl) \gamma_{\alpha} (\psl_2-\ksl_1+\lsl)\gamma^{\mu}(\psl_1-\ksl_1+\lsl)
    \gamma_{\rho'}}{(k_1-k_2)^2 (p_2-k_1+l)^2 (p_1-k_1+l)^2 (p_2-k_2+l)^2 l^2} \Bigr] \non
&\sim & 0,
\label{eq:nlohll33}\\
G^{(1)}_{2i,LL,33} &= & \frac{- e g^4_s}{9} \mathrm{Tr}\Bigl[
\frac{[M_{\rho}\phi^s_{\rho}(x_1)]\gamma^{\alpha} (\ksl_2+\lsl) \gamma^{\rho'} [M_{\rho} \phi^s_{\rho}(x_2)]
\gamma_{\alpha} (\psl_2-\ksl_1+\lsl)\gamma_{\mu}(\psl_1-\ksl_1+\lsl) \gamma_{\rho'}}{(k_1-k_2-l)^2 (p_2-k_1+l)^2
(p_1-k_1+l)^2 (k_2+l)^2 l^2}\Bigr] \non
&\sim & 0,
\label{eq:nloill33}\\
G^{(1)}_{2j,LL,33} &= & \frac{e g^4_s }{9} \mathrm{Tr}\Bigl[
\frac{[M_{\rho}\phi^s_{\rho}(x_1)]\gamma_{\rho'} (\ksl_1-\lsl) \gamma^{\alpha} (\ksl_2-\lsl) \gamma^{\rho'}
[M_{\rho} \phi^s_{\rho}(x_2)]\gamma_{\alpha} (\psl_2-\ksl_1) \gamma_{\mu}}{(k_1-k_2)^2
(p_2-k_1)^2 (k_1-l)^2 (k_2-l)^2 l^2} \Bigr]\non
&\sim & - \frac{1}{8}\Phi^{(1),s}_{\rho, e}(x_1,\xi_1) \otimes G^{(0)}_{a,LL,33}( x_1; x_2),
\label{eq:nlojll33}\\
G^{(1)}_{2k,LL,33} &= & \frac{ - e g^4_s }{9} \mathrm{Tr}\Bigl[
\frac{[M_{\rho}\phi^s_{\rho}(x_1)]\gamma_{\rho'} (\ksl_1-\lsl) \gamma^{\alpha}[M_{\rho} \phi^s_{\rho}(x_2)]
 \gamma^{\rho'}(\psl_2-\ksl_2-\lsl)\gamma_{\alpha}(\psl_2-\ksl_1) \gamma_{\mu}}{(k_1-k_2-l)^2
 (p_2-k_1)^2 (k_1-l)^2 (p_2-k_2-l)^2 l^2}\Bigr]\non
&\sim &\frac{1}{8}\Phi^{(1),s}_{\rho, e}(x_1,\xi_1) \otimes G^{'(0)}_{a,LL,33}( x_1,\xi_1; x_2),
\label{eq:nlokll33}
\eeq}
where the NLO $\rho$ meson wave function $\Phi^{(1),s}_{\rho}$ is of the form
{\small
\beq
\Phi^{(1),s}_{\rho, e}(x_1,\xi_1) = \frac{i g^2_s C_F}{4} \mathrm{Tr} \Bigl[
    \frac{(\ksl_1-\lsl) \gamma^{\rho}  v_{\rho'}}
    {(k_1-l)^2 l^2 (v\cdot l)} \Bigr],
\label{eq:nlofrhos}
\eeq}
which absorbs all the collinear divergences. The new LO hard amplitudes $G^{(0)}_{a,LL,33}$ appeared in
Eqs.~(\ref{eq:nloell33},\ref{eq:nlogll33},\ref{eq:nlojll33},\ref{eq:nlokll33}) are of the following form,
{\small
\beq
G^{(0)}_{a,LL,33}(\xi_1;x_2) &=& -\frac{i e g^2_s C_F}{2} \mathrm{Tr}\Bigl[
\frac{[ M_{\rho} \phi^{s}_{\rho}(x_1)]\gamma^{\alpha}[ M_{\rho} \phi^{s}_{\rho}(x_2)]  \gamma_{\alpha}
(\psl_2 - \ksl_1 + \lsl) \gamma_{\mu}}{(p_2-k_1+l)^2 (k_1-k_2-l)^2}\Bigr],
\label{eq:loall331} \\
G^{(0)}_{a,LL,33}(x_1, \xi_1; x_2) &=& -\frac{i e g^2_s C_F}{2} \mathrm{Tr}\Bigl[
\frac{[ M_{\rho} \phi^{s}_{\rho}(x_1)]\gamma^{\alpha}[ M_{\rho} \phi^{s}_{\rho}(x_2)]
\gamma_{\alpha} (\psl_2 - \ksl_1 + \lsl) \gamma_{\mu}}{(p_2-k_1+l)^2 (k_1-k_2)^2} \Bigr],
\label{eq:loall332} \\
G^{'(0)}_{a,LL,33}(x_1, \xi_1; x_2)&=& -\frac{i e g^2_s C_F}{2} \mathrm{Tr}\Bigl[
\frac{[ M_{\rho} \phi^{s}_{\rho}(x_1)]\gamma^{\alpha}[ M_{\rho} \phi^{s}_{\rho}(x_2)]
\gamma_{\alpha} (\psl_2 - \ksl_1 ) \gamma_{\mu}}{(p_2-k_1)^2 (k_1-k_2-l)^2} \Bigr].
\label{eq:loall333}
\eeq}

Because there is no NLO infrared corrections to LO hard amplitude $G^{(1)}_{a,LL,33}$ from the gluon radiated from the up quark line,
so we can just sum up the infrared contribution generated from the gluon radiated from the down anti-quark line:
{\small
\beq
&&G^{(1)}_{2down,LL,33}(x_1; x_2) = \sum_{n=e,g,j,k}G^{(1)}_{2n,LL,33}(x_1; x_2) \non
  =&& \Phi^{(1)}_{\rho,e}(x_1,\xi_1) \otimes \left[ \frac{7}{16}G^{(0)}_{LL,33}(x_1; x_2)- \frac{1}{16} G^{(0)}_{LL,33}(x_1,\xi_1; x_2)
                                        - \frac{7}{8} G^{'(0)}_{LL,33}(x_1,\xi_1; x_2) \right].
\label{eq:nlodownll33}
\eeq}
It's manifest that the soft divergences generated from Figs.~\ref{fig:fig2}(j,k) only and will be
canceled by each other, while the collinear divergence can also be absorbed into the NLO T3 longitudinal
meson DA $\Phi^{(1),s}_{\rho}$, just like what we have done for the collinear divergences in the NLO corrections
to $G^{(1)}_{a,LL,22}$ above.
The nonlocal hadronic matrix element of $\Phi^{(1),s}_{\rho}$ with the scalar structure $I/2$ can be written as
{\small
\beq
\Phi^{(1),s}_{\rho}&=&\frac{1}{2N_c P^+_1} \int \frac{dy^-}{2\pi} e^{-i x p^+_1 y^-}
    <0| \overline{q}(y^-) \frac{1}{2} (-i g_s)  \int^{y^-}_{0} dz v
\cdot A(zv) q(0)| \rho(p_1)>.
\label{eq:nlo-rhos}
\eeq}
Here only the three sub-diagrams Figs.~\ref{fig:fig3}(c,d,g), where the gluon was radiated from the down anti-quark
line, will contribute.

Following the similar procedures, we can also calculate the NLO infrared corrections to the remaining four
LO hard amplitudes $G^{(0)}_{a,LT,23}(x_1;x_2)$, $G^{(0)}_{a,TL,23}(x_1;x_2)$, $G^{(0)}_{a,TL,32}(x_1;x_2)$
and $G^{(0)}_{a,TT,33}(x_1;x_2)$  as listed  Eqs.~(\ref{eq:lolt23}-\ref{eq:lott33}).
The explicit expressions for the NLO corrections to these four LO amplitudes and their factorization
are all given in detail in Appendix A.

After absorbing the collinear divergences, the NLO transverse $\rho$ meson wave functions can also be
written in the nonlocal matrix element with the gluon momentum along the light cone and flow in a finite
interval $(0,y^-)$:
{\small
\beq
\Phi^{(1),T}_{\rho}&=&\frac{1}{2N_c P^+_1} \int \frac{dy^-}{2\pi} e^{-i x p^+_1 y^-}
    <0| \overline{q}(y^-) \frac{\gamma_{\bot}\gamma^+}{2} (-i g_s)  \int^{y^-}_{0} dz v
\cdot A(zv) q(0)| \rho(p_1)>,
\label{eq:nlo-rhoT}\\
\Phi^{(1),v}_{\rho}&=&\frac{1}{2N_c P^+_1} \int \frac{dy^-}{2\pi} e^{-i x p^+_1 y^-}
    <0| \overline{q}(y^-) \frac{\gamma_{\bot}}{2} (-i g_s)  \int^{y^-}_{0} dz v
\cdot A(zv) q(0)| \rho(p_1)>,
\label{eq:nlo-rhov}\\
\Phi^{(1),a}_{\rho}&=&\frac{1}{2N_c P^+_1} \int \frac{dy^-}{2\pi} e^{-i x p^+_1 y^-}
    <0| \overline{q}(y^-) \frac{\gamma_{\bot}\gamma_5}{2} (-i g_s)  \int^{y^-}_{0} dz v
\cdot A(zv) q(0)| \rho(p_1)>.
\label{eq:nlo-rhoa}
\eeq}
It's reasonable to assume that the NLO transverse $\rho$ meson wave functions in above equations
have the same forms as those we got for the $\rho \gamma^* \to \pi$ transition process \cite{prd90-076001},
which satisfies the universality requirement of the non-perturbative infrared physics.

\begin{figure}[tb]
  \centering
  \vspace{0cm}
  \begin{center}
  \leftline{\epsfxsize=14cm\epsffile{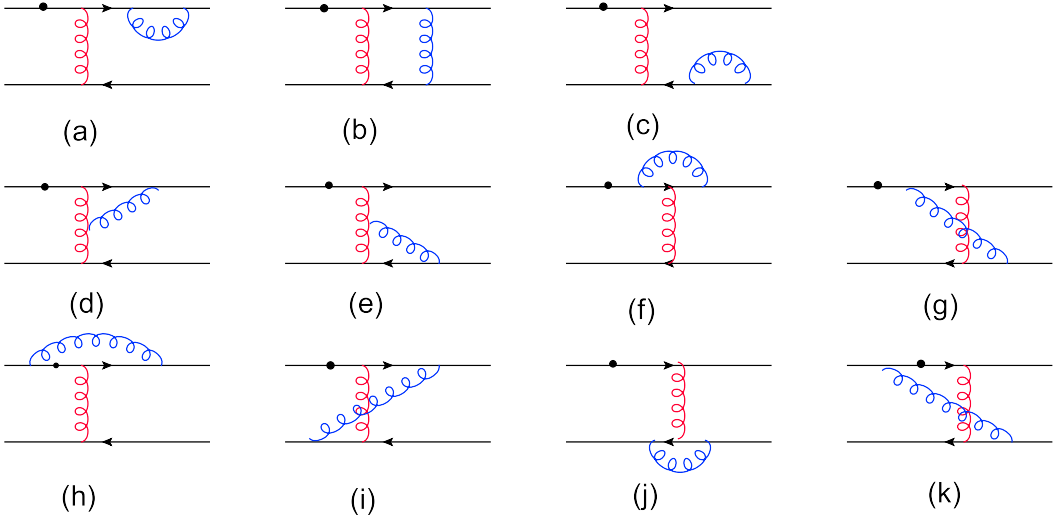}}
  \end{center}
  \vspace{-1cm}
  \caption{Analogous to Fig.~(\ref{fig:fig2}), but with the additional gluon (blue curves) emitted from
  the quark (anti-quark) lines of the final state $\rho$ meson. }
  \label{fig:fig4}
  \end{figure}

Now let's turn our attentions to the infrared contributions in ${\calo}(\alpha^2_s)$) radiative
corrections to Fig.~\ref{fig:fig1}(a) with the gluon emitted from the final $\rho$ meson as shown
explicitly in Fig.~\ref{fig:fig4}, where the infrared divergences would appear when the radiated (blue) gluon
is parallel with the final $\rho$ meson momentum $p_2$.
Why we do this is to show that factorization is still exactly work at the NLO for the gluon radiation from the final state $\rho$ meson,
even though there exist asymmetry for the initial and final $\rho$ meson in Fig.~\ref{fig:fig1}(a)
because of the choice of the weak vector position.
Right here, as an example, we will simply present the detailed NLO corrections for the LO amplitudes $G^{(0)}_{a,LL,22}$ only,
extract the NLO final $\rho$ meson wave function $\Phi^{(1)}_{\rho}(\xi_2,x_2)$,
but do not present the NLO corrections for other LO amplitudes for the sake of the concision of this paper
because the calculations and the factorization processes are very similar.

The NLO amplitudes of reducible sub-diagrams Fig.~\ref{fig:fig4}(a,b,c) are listed as
{\small
\beq
G^{(1)}_{4a,LL,22} &=& -\frac{1}{2} \frac{e g^4_s C^2_F}{2} \mathrm{Tr}\Bigl[
    \frac{[\epsl_{1L} M_{\rho} \phi_{\rho}(x_1)]\gamma^{\alpha} [\epsl_{2L} M_{\rho} \phi_{\rho}(x_2)]\gamma_{\rho'}
    (\psl_2-\ksl_2+\lsl)\gamma^{\rho'} }{(p_2-k_1)^2 (k_1-k_2)^2 (p_2-k_2)^2 (p_2-k_2+l)^2 l^2} \non
&& \cdot  (\psl_2- \ksl_2) \gamma_{\alpha} (\psl_2-\ksl_1) \gamma_{\mu}\Bigr] \non
=&& \frac{1}{2}G^{(0)}_{a,LL,22}(x_1; x_2) \otimes \Phi^{(1)}_{\rho, a}(\xi_2,x_2) ,
\label{eq:nlo5all22}\\
G^{(1)}_{4b,LL,22} &=& \frac{ e g^4_s C^2_F}{2} \mathrm{Tr} \Bigl[
    \frac{[\epsl_{1L} M_{\rho} \phi_{\rho}(x_1)] \gamma^{\alpha}(\ksl_2-\lsl)\gamma_{\rho'} [\epsl_{2L} M_{\rho} \phi_{\rho}(x_2)]
    \gamma_{\rho'}}{(p_2-k_1)^2 (k_1-k_2+l)^2 (p_2-k_2+l)^2 (k_2-l)^2 l^2} \non
&& \cdot (\psl_2-\ksl_2+\lsl) \gamma_{\alpha} (\psl_2-\ksl_1) \gamma_{\mu}\Bigr]\non
&=& G^{(0)}_{a,LL,22}(x_1; x_2)\otimes \Phi^{(1)}_{\rho, b}(\xi_2,x_2),
\label{eq:nlo5bll22}\\
G^{(1)}_{4c,LL,22} &=&- \frac{1}{2} \frac{e g^4_s C^2_F}{2} \mathrm{Tr} \Bigl[
    \frac{[\epsl_{1L} M_{\rho} \phi_{\rho}(x_1)] \gamma^{\alpha}\ksl_2\gamma_{\rho'} (\ksl_2-\lsl)\gamma^{\rho'}[\epsl_{2L} M_{\rho} \phi_{\rho}(x_2)] \gamma_{\alpha} (\psl_2-\ksl_1)\gamma_{\mu} }
    {(p_2-k_1)^2 (k_1-k_2)^2 (k_2-l)^2 (k_2)^2 l^2} \Bigr]\non
=&& \frac{1}{2}G^{(0)}_{a,LL,22}(x_1; x_2) \otimes \Phi^{(1)}_{\rho, c}(\xi_2,x_2),
\label{eq:nlo5cll22}
\eeq }
with the NLO wave functions $\Phi^{(1)}_{\rho, i}, i=a,b,c$
{\small
\beq
\Phi^{(1)}_{\rho, a}(\xi_2,x_2)&=& \frac{- i g^2_s C_F}{4} \mathrm{Tr} \Bigl[
\frac{\gamma^-_{\rho} \gamma^+_{\rho}\gamma_{\rho'} (\psl_2-\ksl_2+\lsl) \gamma^{\rho'}(\psl_2-\ksl_2)
}{(p_2-k_2)^2 (p_2-k_2+l)^2 l^2} \Bigr],  \non
\Phi^{(1)}_{\rho, b}(\xi_2,x_2) &=& \frac{i g^2_s C_F}{4} \mathrm{Tr} \Bigl[
\frac{\gamma^-_{\rho}(\ksl_2-\lsl)\gamma_{\rho'} \gamma^+_{\rho}\gamma_{\rho'}  (\psl_2-\ksl_2+\lsl)}
 {(p_2-k_2+l)^2 (k_2-l)^2 l^2} \Bigr],  \non
\Phi^{(1)}_{\rho, c}(\xi_2,x_2) &=& \frac{- i g^2_s C_F}{4} \mathrm{Tr} \Bigl[
\frac{ \gamma^-_{\rho} \ksl_2\gamma^{\rho'} (\ksl_2-\lsl)\gamma_{\rho'}\gamma^+_{\rho}}
{(k_2-l)^2 (k_2)^2 l^2}\Bigr],
\label{eq:nlorho2-abc}
\eeq }

Amplitudes for the irreducible sub-diagrams Fig.~\ref{fig:fig4}(d-k) are collected with the similar
treatments as for Fig.~\ref{fig:fig2}:
{\small
\beq
G^{(1)}_{4d,LL,22} &=& - e g^4_s  \mathrm{Tr}
\Bigl[ \frac{[\epsl_{1L} M_{\rho} \phi_{\rho}(x_1)] \gamma^{\alpha}[\epsl_{2L} M_{\rho} \phi_{\rho}(x_2)]
    \gamma^{\gamma}(\psl_2-\ksl_2+\lsl)\gamma_{\beta}(\psl_2-\ksl_1) \gamma^{\mu} F_{\alpha \gamma \beta}}
    {(p_2-k_2+l)^2 (p_2-k_1)^2 (k_1-k_2+l)^2 (k_1-k_2)^2 l^2} \Bigr] \non
  &\sim & \frac{9}{8}\left [G^{(0)}_{a,LL,22}(x_1; x_2) - G^{(0)}_{a,LL,22}(x_1; \xi_2)\right]  \otimes \Phi^{(1)}_{\rho, d}(\xi_2,x_2),
\label{eq:nlo5dll22} \\
G^{(1)}_{4e,LL,22} &=&  e g^4_s \mathrm{Tr}
\Bigl[ \frac{[\epsl_{1L} M_{\rho} \phi_{\rho}(x_1)] \gamma^{\alpha}(\ksl_2-\lsl)\gamma^{\gamma}
    [\epsl_{2L} M_{\rho} \phi_{\rho}(x_2)] \gamma^{\beta} (\psl_2-\ksl_1) \gamma_{\mu} F_{\alpha \gamma \beta}}
    {(p_2-k_1)^2 (k_1-k_2)^2 (k_1-k_2+l)^2 (k_2-l)^2 l^2}\Bigr] \non
  &\sim & \frac{9}{8}\left [G^{(0)}_{a,LL,22}(x_1; x_2) - G^{(0)}_{a,LL,22}(x_1; \xi_2) \right] \otimes \Phi^{(1)}_{\rho, e}(\xi_2,x_2),
\label{eq:nlo5ell22}\\
G^{(1)}_{4f,LL,22} &=& \frac{e g^4_s C^2_F }{2} \mathrm{Tr}
\Bigl[ \frac{[\epsl_{1L} M_{\rho} \phi_{\rho}(x_1)] \gamma^{\alpha}[\epsl_{2L} M_{\rho} \phi_{\rho}(x_2)] \gamma^{\rho'}(\psl_2-\ksl_2+\lsl)\gamma_{\alpha} }
    {(p_2-k_1)^2 (k_1-k_2)^2 (p_2-k_1+l)^2 l^2 (p_2-k_2+l)^2} \non
&& \cdot (\psl_2-\ksl_1+\lsl) \gamma_{\rho'}(\psl_2-\ksl_1) \gamma_{\mu}\Bigr]\non
&\sim & -\frac{1}{8}\left [G^{(0)}_{a,LL,22}(x_1; x_2) - G^{(0)}_{a,LL,22}(x_1;x_2, \xi_2)\right]  \otimes  \Phi^{(1)}_{\rho, d}(\xi_2,x_2) ,
\label{eq:nlo5fll22}\\
G^{(1)}_{4g,LL,22} &= &\frac{- e g^4_s C^2_F}{2} \mathrm{Tr} \Bigl[
    \frac{[\epsl_{1L} M_{\rho} \phi_{\rho}(x_1)] \gamma^{\alpha}(\ksl_2-\lsl)\gamma^{\rho'}[\epsl_{2L} M_{\rho} \phi_{\rho}(x_2)]\gamma_{\alpha}}
    {(p_2-k_1)^2 (k_1-k_2+l)^2 (k_2-l)^2 l^2 (p_2-k_1-l)^2} \non
&& \cdot  (\psl_2-\ksl_1-\lsl) \gamma_{\rho'} (\psl_2-\ksl_1)\gamma_{\mu} \Bigr] \non
&\sim & \frac{1}{8}\left [G^{(0)}_{a,LL,22}(x_1; \xi_2) - G^{'(0)}_{a,LL,22}(x_1;x_2, \xi_2)\right]  \otimes  \Phi^{(1)}_{\rho, e}(\xi_2,x_2),
\label{eq:nlo5gll22}\\
G^{(1)}_{4h,LL,22} &=& \frac{ e g^4_s }{9} \mathrm{Tr}
\Bigl[ \frac{[\epsl_{1L} M_{\rho} \phi_{\rho}(x_1)]\gamma^{\alpha}[\epsl_{2L} M_{\rho} \phi_{\rho}(x_2)]
    \gamma^{\rho'}(\psl_2-\ksl_2+\lsl)\gamma_{\alpha} }{(p_2-k_1+l)^2 (k_1-k_2)^2 (p_2-k_2+l)^2 l^2 (p_1-k_1+l)^2 } \non
&& \cdot  (\psl_2-\ksl_1+\lsl) \gamma_{\mu} (\psl_1-\ksl_1+\lsl)\gamma_{\rho'} \Bigr] \non
 & \sim &-\frac{1}{8}  G^{(0)}_{a,LL,22}(x_1; x_2, \xi_2)\otimes\Phi^{(1)}_{\rho, d}(\xi_2,x_2),
\label{eq:nlo5hll22}\\
G^{(1)}_{4i,LL,22} &=& \frac{- e g^4_s }{9} \mathrm{Tr} \Bigl[
\frac{[\epsl_{1L} M_{\rho} \phi_{\rho}(x_1)]\gamma^{\rho'}(\ksl_1+\lsl)\gamma^{\alpha}][\epsl_{2L}
M_{\rho} \phi_{\rho}(x_2)]\gamma_{\rho'}}{(p_2-k_1)^2(k_1-k_2+l)^2 (p_2-k_2+l)^2 l^2 (k_1+l)^2 } \non
&& \cdot (\psl_2-\ksl_2+\lsl) \gamma_{\alpha} (\psl_2-\ksl_1)\gamma_{\mu} \Bigr ]\non
 & \sim & \frac{1}{8} G^{(0)}_{a,LL,22}(x_1;\xi_2) \otimes \Phi^{(1)}_{\rho, d}(\xi_2,x_2),
\label{eq:nlo5ill22}\\
G^{(1)}_{4j,LL,22} &=& \frac{e g^4_s }{9} \mathrm{Tr} \Bigl[
    \frac{[\epsl_{1L} M_{\rho} \phi_{\rho}(x_1)]\gamma_{\rho'} (\ksl_1-\lsl) \gamma^{\alpha}(\ksl_2-\lsl)\gamma^{\rho'}[\epsl_{2L} M_{\rho} \phi_{\rho}(x_2)]\gamma_{\alpha} (\psl_2-\ksl_1) \gamma_{\mu}}
    {(p_2-k_1)^2 (k_1-k_2)^2 (k_2-l)^2 l^2 (k_1-l)^2} \Bigr]\non
&  \sim & -\frac{1}{8} G^{(0)}_{a,LL,22}( x_1; x_2) \otimes \Phi^{(1)}_{\rho, e}(\xi_2,x_2) ,
\label{eq:nlo5jll22}\\
G^{(1)}_{4k,LL,22} &=& \frac{- e g^4_s }{9} \mathrm{Tr} \Bigl[
    \frac{[\epsl_{1L} M_{\rho} \phi_{\rho}(x_1)]\gamma^{\alpha}(\ksl_2-\lsl)\gamma^{\rho'}[\epsl_{2L} M_{\rho} \phi_{\rho}(x_2)]\gamma_{\alpha}}{(p_2-k_1-l)^2 (k_1-k_2+l)^2 (k_2-l)^2 l^2 (p_1-k_1-l)^2} \non
&& \cdot  (\psl_2-\ksl_1-\lsl) \gamma_{\mu} (\psl_1-\ksl_1-\lsl)\gamma_{\rho'} \Bigr] \non
 & \sim &\frac{1}{8} G^{'(0)}_{a,LL,22}( x_1; x_2,\xi_2) \otimes \Phi^{(1)}_{\rho, e}(\xi_2,x_2),
\label{eq:nlo5kll22}
\eeq }
where the NLO final state $\rho$ meson longitudinal wave function $\Phi^{(1)}_{\rho, d},~\Phi^{(1)}_{\rho, e}$ obtained from
the NLO amplitudes $G^{(1)}_{4d,LL,22},~G^{(1)}_{4e,LL,22}$ respectively are of the following form:
{\small
\beq
\Phi^{(1)}_{\rho, d}(\xi_2,x_2) &=& -\frac{ i g^2_s C_F}{4} \mathrm{Tr}
\Bigl[\frac{[\gamma^-_{\rho} \gamma^+_{\rho}]\gamma^{\rho'}(\psl_2-\ksl_2+\lsl)n_{\rho'}}
    {(p_2-k_2+l)^2 l^2 (n\cdot l)} \Bigr].
\label{eq:nlorho2-d}\\
\Phi^{(1)}_{\rho, e}(\xi_2,x_2) &=& \frac{i g^2_s C_F}{4} \mathrm{Tr}
\Bigl[\frac{[\gamma^-_{\rho}\gamma^+_{\rho}](\ksl_2-\lsl)\gamma^{\rho'}  n_{\rho'}}
    {(k_2-l)^2 l^2 (n\cdot l)}\Bigr],
\label{eq:nlorho2-e}
\eeq}
and the modified LO hard kernels with the gluon momentum flow, partly flowing into the original LO
hard kernel $G^{(0)}_{a,LL,22}(x_1;x_2)$, as in the case for the NLO correction from the initial meson
radiations, are defined as,
{\small
\beq
G^{(0)}_{a,LL,22}(x_1;\xi_2)&=&-\frac{i e g^2_s C_F}{2} \mathrm{Tr}
\Bigl[ \frac{[\epsl_{1L} M_{\rho} \phi_{\rho}(x_1)] \gamma^{\alpha} [\epsl_{2L} M_{\rho} \phi_{\rho}(x_2)]\gamma_{\alpha}(\psl_2 - \ksl_1) \gamma_{\mu}}
    {(p_2-k_1)^2 (k_1-k_2+l)^2} \Bigr],
\label{eq:loall2211} \\
G^{(0)}_{a,LL,22}(x_1;x_2, \xi_2 )&=&-\frac{i e g^2_s C_F}{2} \mathrm{Tr}
\Bigl[  \frac{[\epsl_{1L} M_{\rho} \phi_{\rho}(x_1)] \gamma^{\alpha} [\epsl_{2L} M_{\rho} \phi_{\rho}(x_2)]\gamma_{\alpha}(\psl_2 - \ksl_1 ) \gamma_{\mu}}
    {(p_2-k_1+l)^2 (k_1-k_2)^2} \Bigr],
\label{eq:loall2222} \\
G^{'(0)}_{a,LL,22}(x_1;  x_2,\xi_2)&=&-\frac{i e g^2_s C_F}{2} \mathrm{Tr}
\Bigl[ \frac{[\epsl_{1L} M_{\rho} \phi_{\rho}(x_1)] \gamma^{\alpha} [\epsl_{2L} M_{\rho} \phi_{\rho}(x_2)]\gamma_{\alpha}(\psl_2 - \ksl_1 ) \gamma_{\mu}}
    {(p_2-k_1-l)^2 (k_1-k_2+l)^2} \Bigr].
\label{eq:loall2233}
\eeq}

Summing up the infrared amplitudes sorted by the gluon radiated either from the right-up (Fig.~\ref{fig:fig4}(d,f,h,i))
or from the right-down (Fig.~\ref{fig:fig4}(e,g,j,k)) quark line,
we obtain their total factorization formulation for the final state gluon corrections to the LO amplitudes
$G^{(0)}_{a,LL,22}$:
{\small
\beq
G^{(1)}_{4up,LL,22} (x_1; x_2)&=& \sum_{m=d,f,h,i}G^{(1)}_{4m,LL,22}(x_1; x_2) \non
  &=&  \frac{7}{8}\left[G^{(0)}_{a,LL,22}(x_1; x_2) - G^{(0)}_{a,LL,22}(x_1; \xi_2)\right] \otimes \Phi^{(1)}_{\rho,d}(\xi_2,x_2).
  \label{eq:nloupll22-f}\\
G^{(1)}_{4down,LL,22}(x_1; x_2) &=& \sum_{n=e,g,j,k}G^{(1)}_{4n,LL,22}(x_1; x_2) \non
  &=& \frac{7}{8} \left[G^{(0)}_{a,LL,22}(x_1; x_2)- G^{(0)}_{a,LL,22}(x_1; \xi_2)\right] \otimes \Phi^{(1)}_{\rho,e}(\xi_2,x_2).
\label{eq:nlodownll22-f}
\eeq}

The NLO corrections to the LO amplitude $G^{(0)}_{a,LL,33}$ from the final state radiations can be treated in the same way.
So far the factorizations of Fig.~\ref{fig:fig1}(a) at the NLO for both the initial- and the final state radiation
are explicitly shown above, and we can also write down the longitudinal NLO final $\rho$ meson wave function
$\Phi^{(1)}_{\rho},~\Phi^{(1),s}_{\rho}$
in the similar formula as the NLO initial meson wave function in Eqs.~(\ref{eq:nlo-rho},\ref{eq:nlofrhos}).

\subsection{$k_T$ factorization of the NLO corrections to $\rho \gamma^{\star} \to \rho$}

In order to eliminate the end-point singularity which may be appeared in the small $x$ region under collinear factorization framework,
the $\kt$ factorization frame is developed by picking up previously dropped transversal momentum $\kt$ for the external quark lines.
And a Sudakov factor $e^{-S(t)}$,
which indicating the probability there is no gluon radiation in the localized phase space for exclusive final products,
is emerged with the resummation of the large divergences logarithms and furthermore,
could suppress the soft-dynamics effectively \cite{ppnp51-85}.

In the PQCD approach, because the Sudakov factor increase transversal momentum $k_T$ and longitudinal momentum fraction $x$ simultaneously,
the transversal momentum $k_T$ in the initial and final meson bound state remain far less than the energy scales in the LO hard kernel,
i.e., $k^2_{\rm iT} \ll k_1\cdot k_2$.
This argument apparently support our expectation that the eikonal approximation used in the collinear
factorization are also valid for the $\kt$ factorization.
The collinear factorization operating for $\rho \gamma^{\star} \to \rho$  in the previous  subsection
can be extended into the $k_T$ factorization directly.
The only difference is that we should take care of the transverse momentum $\mathbf{l}_T$ in the
denominator of the gluon propagators.
Of course, we can also retrieve the Feynman-rule of the Wilson line by including $\mathbf{l}_T$
through the Fourier transformation for the gauge field from $A(zv)$ to $\widetilde{A}(l)$
for those NLO wave functions in Eqs.~(\ref{eq:nlo-rho},\ref{eq:nlo-rhos},\ref{eq:nlo-rhoT},\ref{eq:nlo-rhov},
\ref{eq:nlo-rhoa}):
\beq
\int^{\infty}_{0} dz v\cdot A(zv) & \to & \int^{\infty}_{0} dz\int dl ~e^{iz(v \cdot l + i\epsilon)}~v\cdot \widetilde{A}(l)
= i \int dl ~\frac{v_\rho}{v\cdot l} \widetilde{A}^\rho(l),
\label{eq:feyn-wilsonline1} \\
\int^{y^-}_{\infty} dz v\cdot A(zv)
& \to & \int^{y^-}_{0} dz \int dl ~e^{[iz(v \cdot l + i\epsilon)-i\mathbf{l}_T \cdot \mathbf{b}]}~v\cdot \widetilde{A}(l)\non
&=& -i \int dl ~\frac{v_\rho}{v\cdot l} ~e^{[il^+y^- - i\mathbf{l}_T \cdot \mathbf{b}]}~\widetilde{A}^\rho(l),
\label{eq:feyn-220}
\eeq
where the factor $\exp[il^+y^-]$ in the integration in Eq.~(\ref{eq:feyn-220}) will generate the delta function
$\delta(\xi_1-x_1+\frac{l^+}{p_1})$, which imply that the gluon momentum is flowing into the LO hard kernel.
The another factor $\exp[- i\mathbf{l}_T \cdot \mathbf{b}]$ represent the transversal momentum flowing into the LO hard kernel,
accompanied with the function $\delta(\xi_1-x_1+\frac{l^+}{p_1})$.
And this modification can be understood graphically as the deviation of the second half integration for Wilson line
by a transversal interval $\mathbf{b}$ from the light cone direction, as illustrated by
Figs.~\ref{fig:fig4}(a) and \ref{fig:fig4}(b).

\begin{figure}[tb]
\vspace{-2cm}
\begin{center}
\leftline{\epsfxsize=16cm\epsffile{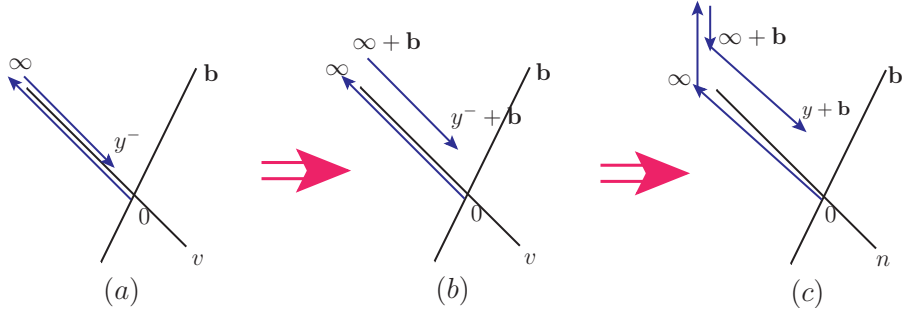}}
\end{center}
\vspace{-17.5cm}
\caption{The deviation of the integral (Wilson link) from the light corn by $\mathbf{b}$ in
the coordinate space for the two-parton meson wave function.}
\label{fig:fig5}
\end{figure}

It's clear to see that the Wilson lines, no matter along the light cone as shown in Fig.~\ref{fig:fig4}(a)
or parallel to the light cone as illustrated in Fig.~\ref{fig:fig4}(b), will generate light cone singularity.
The simple method to resolve this problem is to move the Wilson line a little away from
the light cone, as depicted in Fig.~\ref{fig:fig4}(c), and regularize the non-physical light cone
singularity by the logarithm of $n^2$ with $n^2 \neq 0$.
Even the scheme-dependence induced by this variation has been resumed in Ref.~\cite{li1302}
and found that such effect is small in size.
One problem resulted from such operation is the appearance of the pinched infrared divergence
in the self-energy correction of the Wilson line.
In order to avoid such  kind of infrared divergence,  the transverse momentum dependent distribution (TMD) with
a soft substraction was proposed in Ref.~\cite{ijmpcs4-85}, and another new definition with the choice
of two orthogonal gauge vectors for off light cone Wilson links  also be discussed in Ref.~\cite{jhep1506-013}.
But such topic is apparently beyond the scope of this paper.
In this paper, we are aim to factorize the infrared NLO corrections to those LO amplitudes,
so we still use the variation method to avoid the light cone singularity.

To finish this section, we write down the NLO $\rho$ meson functions obtained from $\rho \gamma^{\star} \to \rho$
at leading and sub-leading twist level in $k_T$ factorization as the following form,
\beq
  \Phi^{(1),T}_{\rho}(x_1,\xi_1;\mathbf{b_1})&=&\frac{1}{2N_c P^+_1} \int \frac{dy^-}{2\pi} \frac{d\mathbf{b_1}}{(2\pi)^2}
  e^{-i x p^+_1 y^- + i \mathbf{k_{1T}} \cdot \mathbf{b_1}} \non
  &&\cdot <0| \overline{q}(y^-) \frac{\gamma^b_{\bot} \gamma^+}{4} (-i g_s) \int^{y}_{0} dz n \cdot A(zn) q(0)|\rho(p_1)>,
  \label{eq:nlorhoTkt}
  \eeq
  \beq
  \Phi^{(1),v}_{\rho}(x_1,\xi_1;\mathbf{b_1})=&&\frac{1}{2N_c P^+_1} \int \frac{dy^-}{2\pi} \frac{d\mathbf{b_1}}{(2\pi)^2}
  e^{-i x p^+_1 y^- + i \mathbf{k_{1T}} \cdot \mathbf{b_1}} \non
  &&\cdot <0|\overline{q}(y^-) \frac{\gamma_{\bot}}{2} (-i g_s)  \int^{y}_{0} dz n \cdot A(zn) q(0)|\rho(p_1)>,
  \label{eq:nlorhovkt}
  \eeq
  \beq
  \Phi^{(1),a}_{\rho}(x_1,\xi_1;\mathbf{b_1})=&&\frac{1}{2N_c P^+_1} \int \frac{dy^-}{2\pi} \frac{d\mathbf{b_1}}{(2\pi)^2}
  e^{-i x p^+_1 y^- + i \mathbf{k_{1T}} \cdot \mathbf{b_1}} \non
  &&\cdot <0|\overline{q}(y^-) \frac{\gamma_5 \gamma_{\bot}}{2} (-i g_s) \int^{y}_{0} dz n \cdot A(zn) q(0)|\rho(p_1)>;
\label{eq:nlorhoakt}
\eeq
\beq
\Phi^{(1)}_{\rho}(x_1,\xi_1;\mathbf{b_1})=&&\frac{1}{2N_c P^+_1} \int \frac{dy^-}{2\pi} \frac{d\mathbf{b_1}}{(2\pi)^2}
  e^{-i x p^+_1 y^- + i \mathbf{k_{1T}} \cdot \mathbf{b_1}} \non
  &&\cdot <0|\overline{q}(y^-) \frac{ \gamma^-}{2} (-i g_s) \int^{y}_{0} dz n \cdot A(zn) q(0)|\rho(p_1)>;
  \label{eq:nlorhokt}
\eeq
\beq
\Phi^{(1),s}_{\rho}(x_1,\xi_1;\mathbf{b_1})=&&\frac{1}{2N_c P^+_1} \int \frac{dy^-}{2\pi} \frac{d\mathbf{b_1}}{(2\pi)^2}
  e^{-i x p^+_1 y^- + i \mathbf{k_{1T}} \cdot \mathbf{b_1}} \non
  &&\cdot <0|\overline{q}(y^-) \frac{1}{2} (-i g_s) \int^{y}_{0} dz n \cdot A(zn) q(0)|\rho(p_1)>.
  \label{eq:nlorhoskt}
  \eeq
With these gauge invariant NLO wave functions, we can then calculate the NLO hard corrections
to $\rho \gamma^{\star} \to \rho, B \to \rho$ transition processes with the factorization hypothesis.

\section{Summarey}

In this paper we firstly investigated the $\rho$ meson electromagnetic transition process and wrote down
it's hard amplitudes according to the polarizations and the light cone expansion power of the initial and
final meson wave functions.
Next we made an explicit demonstration for the collinear factorization of the NLO corrections to the LO amplitudes
for the Feynman diagrams with a gluon radiated from the initial $\rho$ meson.
With the successful separation in collinear factorization, we then extend these results to the case of
the $\kt$ factorization.

From our analytical evaluations we found the following points:
\ben
\item[(a)]
For the considered $\rho \gamma^* \to \rho$ transition process, there are six LO amplitudes with the fixed polarization
and the twist, say $G^{(0)}_{X,IJ,kl}$ with $X=(a,b,c,d)$, $I,J=(L,T)$ (polarization) and $k,l=(2,3)$ (twist),
for each sub-diagram  $``X"$ in Fig.~\ref{fig:fig1}.
Between the four sub-diagrams in Fig.~\ref{fig:fig1} there exists a simple kinematic exchanging symmetry.
Taking Fig.~\ref{fig:fig1}(a) as an example, its LO hard amplitudes have been sorted in terms of polarization
and twist of the meson wave functions and  given in
Eqs.~(\ref{eq:loll22},\ref{eq:loll33},\ref{eq:lolt23},\ref{eq:lotl23},\ref{eq:lotl32},\ref{eq:lott33}).

\item[(b)]
We show explicitly that the factorization hypothesis is valid at the NLO level for $\rho \gamma^{\star} \to \rho$ transition
process, even if we include the transversal momentum of the light external quark lines.
At the NLO level, the soft divergence from different sub-diagrams will cancel each other in the quark level
and the remaining collinear divergences can be absorbed into the NLO meson wave functions $\Phi^{(1)}_{\rho}$.

\item[(c)]
The NLO $\rho$ meson wave functions, which absorb the collinear divergences, can be written in a nonlocal two
quarks hadron matrix elements as given in Eqs.(36,50-53,56-60),
in which the integration path of the gauge factor represent the direction of the collinear gluon momentum.

\een

\begin{acknowledgments}

The authors would like to thank H.N.~Li and C.D.~Lu for long term collaborations and valuable discussions.
Z.J.~Xiao and Y.L.~Zhang are supported by the National Natural Science Foundation of China under Grant No. 11235005
and the Project on Graduate Students¡¯ Education and Innovation of Jiangsu Province under Grant No. KYZZ15-0212.
S.~Cheng acknowledges the support by the DFG Research Unit FOR 1873 "Quark Flavour Physics and Effective Theories".

\end{acknowledgments}


\begin{appendix}

\section{Factorization of the NLO  amplitudes }\label{sec:da}

In this Appendix we supply the explicit expressions for the NLO corrections to the four LO amplitudes
$G^{(0)}_{a,LT,23}$, $G^{(0)}_{a,TL,23}$,  $G^{(0)}_{a,TL,32}$ and $G^{(0)}_{a,TT,33}$ and the relevant
factorization. The infrared divergences are absorbed into the NLO wave functions $\Phi^{(1)}_{\rho, i}$.

\subsection{The NLO amplitudes for $G^{(0)}_{a,LT,23}$}

For all the eleven sub-diagrams in Fig.~\ref{fig:fig2}, the corresponding NLO amplitudes for $G^{(0)}_{a,LT,23}$
are the following
{\small
\beq
G^{(1)}_{2a,LT,23} &=& -\frac{1}{2} \frac{e g^4_s C^2_F}{2} \mathrm{Tr} \Bigl [
    \frac{\gamma^{\alpha}(\psl_2 - \ksl_1)\gamma_{\mu}(\psl_1 - \ksl_1)\gamma^{\rho'} (\psl_1 - \ksl_1 + \lsl) \gamma_{\rho'}[\epsl_{1L} M_{\rho} \phi_{\rho}(x_1)]\gamma^{\alpha}}{(p_2-k_1)^2 (k_1-k_2)^2 (p_1-k_1)^2 (p_1-k_1+l)^2 l^2} \non
 &&   \cdot[\epsl_{2T} M_{\rho} \phi^{v}_{\rho}(x_2)+ M_{\rho} i \epsilon_{\mu'\nu\rho\sigma} \gamma_5 \gamma^{\mu'} \epsilon^{\nu}_{2T} v^{\rho} n^{\sigma} \phi^{a}_{\rho}(x_2)]
    \Bigr] \non
&=& \frac{1}{2} \Phi^{(1)}_{\rho, a} \otimes G^{(0)}_{a,LT,23}(x_1; x_2),
\label{eq:nloalt23}
\eeq
\beq
G^{(1)}_{2b,LT,23} &=& \frac{ e g^4_s C^2_F}{2} \mathrm{Tr}\Bigl [
 \frac{\gamma_{\alpha}(\psl_2 - \ksl_1 + \lsl)\gamma_{\mu} (\psl_1 - \ksl_1 + \lsl) \gamma^{\rho'}[\epsl_{1L} M_{\rho} \phi_{\rho}(x_1)]\gamma^{\rho'} (\ksl_1 - \lsl) \gamma^{\alpha} }{(p_2-k_1+l)^2 (k_1-k_2-l)^2 (p_1-k_1+l)^2 (k_1-l)^2 l^2} \non
  && \cdot [\epsl_{2T} M_{\rho} \phi^{v}_{\rho}(x_2)+ M_{\rho} i \epsilon_{\mu'\nu\rho\sigma} \gamma_5 \gamma^{\mu'} \epsilon^{\nu}_{2T} v^{\rho} n^{\sigma} \phi^{a}_{\rho}(x_2)]
   \Bigr] \non
&=& \Phi^{(1)}_{\rho, b} \otimes G^{(0)}_{a,LT,23}(\xi_1, x_2),
\label{eq:nloblt23}\\
G^{(1)}_{2c,LT,23} &=&- \frac{1}{2} \frac{e g^4_s C^2_F}{2} \mathrm{Tr} \Bigl[
    \frac{\gamma_{\alpha} (\psl_2-\ksl_1)\gamma_{\mu} [\epsl_{1L} M_{\rho} \phi_{\rho}(x_1)] \gamma^{\rho'} (\ksl_1-\lsl)\gamma_{\rho'}\ksl_1\gamma^{\alpha} } {(p_2-k_1)^2 (k_1-k_2)^2 (k_1-l)^2 (k_1)^2 l^2} \non
    \cdot&&[\epsl_{2T} M_{\rho} \phi^{v}_{\rho}(x_2)+ M_{\rho} i \epsilon_{\mu'\nu\rho\sigma} \gamma_5 \gamma^{\mu'} \epsilon^{\nu}_{2T} v^{\rho} n^{\sigma} \phi^{a}_{\rho}(x_2)]
    \Bigr] \non
&=& \frac{1}{2}\Phi^{(1)}_{\rho, c}\otimes G^{(0)}_{a,LT,23}(x_1; x_2)  ,
\label{eq:nloclt23}\\
G^{(1)}_{2d,LT,23} &=& e g^4_s \mathrm{Tr}\Bigl[
    \frac{\gamma_{\beta} (\psl_2 - \ksl_1+ \lsl) \gamma^{\mu}(\psl_1 - \ksl_1 + \lsl)\gamma^{\gamma} F_{\alpha \beta \gamma}[\epsl_{1L} M_{\rho} \phi_{\rho}(x_1)] \gamma^{\alpha} }{(p_2-k_1+l)^2 (k_1-k_2)^2 (p_1-k_1+l)^2 (k_1-k_2-l)^2 l^2} \non
&&    \cdot [\epsl_{2T} M_{\rho} \phi^{v}_{\rho}(x_2)+ M_{\rho} i \epsilon_{\mu'\nu\rho\sigma} \gamma_5 \gamma^{\mu'} \epsilon^{\nu}_{2T} v^{\rho} n^{\sigma} \phi^{a}_{\rho}(x_2)]
    \Bigr] \non
&\sim & \frac{9}{16} \Phi^{(1)}_{\rho, d} \otimes [G^{(0)}_{a,LT,23}(x_1;\xi_1; x_2) - G^{(0)}_{a,LT,23}(\xi_1; x_2)],
\label{eq:nlodlt23}\\
G^{(1)}_{2e,LT,23} &= & - e g^4_s \mathrm{Tr}\Bigl [
    \frac{\gamma_{\beta} (\psl_2 - \ksl_1)\gamma^{\mu} F_{\alpha \beta \gamma}[\epsl_{1L}
M_{\rho} \phi_{\rho}(x_1)] \gamma^{\gamma} (\ksl_1 - \lsl) \gamma^{\alpha}}{(p_2-k_1)^2
(k_1-k_2)^2 (k_1-l)^2 (k_1-k_2-l)^2 l^2} \non
&&     \cdot[\epsl_{2T} M_{\rho} \phi^{v}_{\rho}(x_2)+ M_{\rho} i \epsilon_{\mu'\nu\rho\sigma}
\gamma_5 \gamma^{\mu'} \epsilon^{\nu}_{2T} v^{\rho} n^{\sigma} \phi^{a}_{\rho}(x_2)]
     \Bigr ] \non
&\sim & \frac{9}{16} \Phi^{(1)}_{\rho, e} \otimes [G^{(0)}_{a,LT,23}(x_1; x_2) - G^{'(0)}_{a,LT,23}(x_1;\xi_1;x_2)],
\label{eq:nloelt23}\\
G^{(1)}_{2f,LT,23} &= & \frac{e g^4_s }{9} \mathrm{Tr} \Bigl[
\frac{\gamma_{\alpha} (\psl_2 - \ksl_1) \gamma^{\rho'}(\psl_2-\ksl_1+\lsl) \gamma_{\mu}(\psl_1-\ksl_1+\lsl) \gamma_{\rho'}[\epsl_{1L} M_{\rho} \phi_{\rho}(x_1)]\gamma^{\alpha} }{(p_2-k_1)^2 (k_1-k_2)^2 (p_1-k_1+l)^2 (p_2-k_1+l)^2 l^2} \non
&& \cdot[\epsl_{2T} M_{\rho} \phi^{v}_{\rho}(x_2)+ M_{\rho} i \epsilon_{\mu'\nu\rho\sigma} \gamma_5 \gamma^{\mu'} \epsilon^{\nu}_{2T} v^{\rho} n^{\sigma} \phi^{a}_{\rho}(x_2)]\Bigr]\non
&\sim &  \Phi^{(1)}_{\rho, d} \otimes [G^{(0)}_{a,LT,23}(x_1; x_2) - G^{(0)}_{a,LT,23}(x_1;\xi_1;x_2)] ,
\label{eq:nloflt23}\\
G^{(1)}_{2g,LT,23} &= & \frac{ - e g^4_s}{9} \mathrm{Tr} \Bigl[
\frac{\gamma_{\alpha} (\psl_2 - \ksl_1+\lsl)\gamma^{\rho'} (\psl_2-\ksl_1) \gamma_{\mu}[\epsl_{1L} M_{\rho} \phi_{\rho}(x_1)]\gamma_{\rho'}
(\ksl_1 - \lsl) \gamma^{\alpha} }{(p_2-k_1)^2 (k_1-k_2-l)^2 (p_2-k_1+l)^2 (k_1-l)^2 l^2} \non
&& \cdot[\epsl_{2T} M_{\rho} \phi^{v}_{\rho}(x_2)+ M_{\rho} i \epsilon_{\mu'\nu\rho\sigma} \gamma_5 \gamma^{\mu'}
\epsilon^{\nu}_{2T} v^{\rho} n^{\sigma} \phi^{a}_{\rho}(x_2)]\Bigr]\non
&\sim &  \Phi^{(1)}_{\rho, e} \otimes [G^{'(0)}_{a,LT,23}(x_1;\xi_1; x_2) - G^{(0)}_{a,LT,23}(\xi_1;x_2)] .
\label{eq:nloglt23}\\
G^{(1)}_{2h,LT,23} &= & \frac{e g^4_s }{9} \mathrm{Tr}\Bigl[
    \frac{\gamma^{\rho'} (\psl_2-\ksl_2+\lsl)\gamma_{\alpha} (\psl_2-\ksl_1+\lsl)\gamma^{\mu} (\psl_1-\ksl_1+\lsl)
    \gamma_{\rho'}[\epsl_{1L} M_{\rho} \phi_{\rho}(x_1)]\gamma^{\alpha} }{(k_1-k_2)^2 (p_2-k_1+l)^2 (p_1-k_1+l)^2 (p_2-k_2+l)^2 l^2} \non
&&   \cdot [\epsl_{2T} M_{\rho} \phi^{v}_{\rho}(x_2)+ M_{\rho} i \epsilon_{\mu'\nu\rho\sigma} \gamma_5 \gamma^{\mu'} \epsilon^{\nu}_{2T} v^{\rho}
   n^{\sigma} \phi^{a}_{\rho}(x_2)]\Bigr]\non
&\sim &-\frac{1}{8} \Phi^{(1)}_{\rho, d} \otimes G^{(0)}_{a,LT,23}(x_1;\xi_1;x_2) ,
\label{eq:nlohlt23}\\
G^{(1)}_{2i,LT,23} &= & \frac{- e g^4_s }{9} \mathrm{Tr}\Bigl[
    \frac{\gamma_{\alpha} (\psl_2-\ksl_1+\lsl)\gamma_{\mu} (\psl_1-\ksl_1+\lsl) \gamma_{\rho'}[\epsl_{1L} M_{\rho} \phi_{\rho}(x_1)]\gamma^{\alpha} (\ksl_2+\lsl) \gamma^{\rho'} }{(k_1-k_2-l)^2 (p_2-k_1+l)^2 (p_1-k_1+l)^2 (k_2+l)^2 l^2} \non
&&    \cdot[\epsl_{2T} M_{\rho} \phi^{v}_{\rho}(x_2)+ M_{\rho} i \epsilon_{\mu'\nu\rho\sigma} \gamma_5 \gamma^{\mu'} \epsilon^{\nu}_{2T} v^{\rho} n^{\sigma} \phi^{a}_{\rho}(x_2)]
    \Bigr]\non
&\sim &\frac{1}{8} \Phi^{(1)}_{\rho, d} \otimes G^{(0)}_{a,LT,23}(\xi_1;x_2),
\label{eq:nloilt23}
\eeq
\beq
G^{(1)}_{2j,LT,23} &= & \frac{e g^4_s}{9}  \mathrm{Tr}\Bigl [
\frac{\gamma_{\alpha} (\psl_2-\ksl_1) \gamma_{ \mu}[\epsl_{1L} M_{\rho} \phi_{\rho}(x_1)]\gamma_{\rho'} (\ksl_1-\lsl) \gamma^{\alpha}
(\ksl_2-\lsl) \gamma^{\rho'}}{(k_1-k_2)^2 (p_2-k_1)^2 (k_1-l)^2 (k_2-l)^2 l^2} \non
&&\cdot[\epsl_{2T} M_{\rho} \phi^{v}_{\rho}(x_2)+ M_{\rho} i \epsilon_{\mu'\nu\rho\sigma} \gamma_5 \gamma^{\mu'}
\epsilon^{\nu}_{2T} v^{\rho} n^{\sigma} \phi^{a}_{\rho}(x_2)]\Bigr] \non
&\sim & -\frac{1}{8} \Phi^{(1)}_{\rho, e} \otimes G^{(0)}_{a,LT,23}(x_1;x_2),
\label{eq:nlojlt23}\\
G^{(1)}_{2k,LT,23} &= & \frac{- e g^4_s }{9}  \mathrm{Tr} \Bigl [
\frac{\gamma^{\rho'}(\psl_2-\ksl_2-\lsl)\gamma_{\alpha}(\psl_2-\ksl) \gamma_{\mu}[\epsl_{1L} M_{\rho} \phi_{\rho}(x_1)]\gamma_{\rho'} (\ksl_1-\lsl) \gamma^{\alpha}}{(k_1-k_2-l)^2 (p_2-k_1)^2 (k_1-l)^2 (p_2-k_2-l)^2 l^2} \non
&& \cdot[\epsl_{2T} M_{\rho} \phi^{v}_{\rho}(x_2)+ M_{\rho} i \epsilon_{\mu'\nu\rho\sigma} \gamma_5 \gamma^{\mu'} \epsilon^{\nu}_{2T} v^{\rho} n^{\sigma}
\phi^{a}_{\rho}(x_2)] \Bigr] \non
&\sim & \frac{1}{8} \Phi^{(1)}_{\rho, e} \otimes G^{'(0)}_{a,LT,23}(x_1;\xi_1;x_2),
\label{eq:nloklt23}
\eeq}
where the NLO $\rho$ meson wave function $\Phi^{(1)}_{\rho}$ have the same expressions as those for
the NLO corrections to the amplitude $G^{(0)}_{a,LL,22}$ in Eqs.~(\ref{eq:nlorho-abc},\ref{eq:nlorho-d},\ref{eq:nlorho-e}),
and the new LO hard kernels appeared in above equations are collected as follows,
{\small
\beq
G^{(0)}_{a,LT,23}(\xi_1;x_2) &=& -\frac{i e g^2_s C_F}{2} \mathrm{Tr} \Bigl[
\frac{[\epsl_{2T} M_{\rho} \phi^{v}_{\rho}(x_2)+ M_{\rho} i \epsilon_{\mu'\nu\rho\sigma} \gamma_5 \gamma^{\mu'} \epsilon^{\nu}_{2T} v^{\rho} n^{\sigma} \phi^{a}_{\rho}(x_2)]}{(p_2-k_1+l)^2
(k_1-k_2-l)^2} \non
&&\cdot  \gamma_{\alpha} (\psl_2 - \ksl_1 + \lsl) \gamma_{\mu}[\epsl_{1L} M_{\rho} \phi_{\rho}(x_1)]\gamma^{\alpha} \Bigr],
\label{eq:lolt231}\\
G^{(0)}_{a,LT,23}(x_1,\xi_1;x_2) &=& -\frac{i e g^2_s C_F}{2} \mathrm{Tr} \Bigl[
\frac{[\epsl_{2T} M_{\rho} \phi^{v}_{\rho}(x_2)+ M_{\rho} i \epsilon_{\mu'\nu\rho\sigma} \gamma_5 \gamma^{\mu'} \epsilon^{\nu}_{2T} v^{\rho} n^{\sigma} \phi^{a}_{\rho}(x_2)]}{(p_2-k_1+l)^2
(k_1-k_2)^2}\non
&&\cdot  \gamma_{\alpha} (\psl_2 - \ksl_1 + \lsl) \gamma_{\mu}[\epsl_{1L} M_{\rho} \phi_{\rho}(x_1)]\gamma^{\alpha} \Bigr] ,
\label{eq:lolt232}\\
G^{'(0)}_{a,LT,23}(x_1,\xi_1;x_2) &=& -\frac{i e g^2_s C_F}{2} \mathrm{Tr} \Bigl[
\frac{[\epsl_{2T} M_{\rho} \phi^{v}_{\rho}(x_2)+ M_{\rho} i \epsilon_{\mu'\nu\rho\sigma} \gamma_5 \gamma^{\mu'} \epsilon^{\nu}_{2T} v^{\rho} n^{\sigma} \phi^{a}_{\rho}(x_2)]}{(p_2-k_1)^2 (k_1-k_2-l)^2}\non
&&\cdot  \gamma_{\alpha} (\psl_2 - \ksl_1 ) \gamma_{\mu}[\epsl_{1L} M_{\rho} \phi_{\rho}(x_1)]\gamma^{\alpha}\Bigr].
\label{eq:lolt233}
\eeq}

\subsection{the NLO amplitudes for $G^{(0)}_{a,TL,23}$}

The first three NLO amplitudes $G^{(1)}_{X,TL,23}$ for sub-diagrams Figs.~\ref{fig:fig2}(a,b,c) are the following
{\small
\beq
G^{(1)}_{2a,TL,23} &=& \frac{1}{2} \frac{-e g^4_s C^2_F}{2} \mathrm{Tr} \Bigl [
    \frac{(\psl_1 - \ksl_1)\gamma^{\rho'} (\psl_1 - \ksl_1 + \lsl) \gamma_{\rho'}[\epsl_{1T} \psl_1\phi^T_{\rho}(x_1)]\gamma^{\alpha}[M_{\rho}\phi^s_{\rho}(x_2)]\gamma_{\alpha}(\psl_2 - \ksl_1)\gamma_{\mu}}
    {(p_2-k_1)^2 (k_1-k_2)^2 (p_1-k_1)^2 (p_1-k_1+l)^2 l^2} \Bigr]\non
&=& \frac{1}{2} \Phi^{(1),T}_{\rho, a} \otimes G^{(0)}_{a,TL,23}(x_1; x_2),
\label{eq:nloatl23}\\
G^{(1)}_{2b,TL,23} &=& \frac{ e g^4_s C^2_F}{2}\mathrm{Tr} \Bigl [
    \frac{[\epsl_{1T} M_{\rho} \phi^v_{\rho}(x_1)]\gamma^{\rho'} (\ksl_1 - \lsl) \gamma^{\alpha} [ M_{\rho} \phi^s_{\rho}(x_2)] \gamma_{\alpha}
    (\psl_2 - \ksl_1 + \lsl)\gamma_{\mu}(\psl_1 - \ksl_1 + \lsl) \gamma^{\rho'}}{(p_2-k_1+l)^2 (k_1-k_2-l)^2 (p_1-k_1+l)^2 (k_1-l)^2 l^2}
    \Bigr]  \non
&=& \Phi^{(1),T}_{\rho, b} \otimes G^{(0)}_{a,TL,23}(\xi_1; x_2),
\label{eq:nlobtl23}\\
G^{(1)}_{2c,TL,23} &=& \frac{1}{2} \frac{-e g^4_s C^2_F}{2}\mathrm{Tr}\Bigl [
    \frac{[\epsl_{1T}\psl_1 \phi^T_{\rho}(x_1)]\gamma^{\rho'}(\ksl_1 - \lsl) \gamma_{\rho'} \ksl_1 \gamma^{\alpha} [M_{\rho} \phi^s_{\rho}(x_2)] \gamma_{\alpha}
    (\psl_2 - \ksl_1) \gamma_{\mu}}{(p_2-k_1)^2 (k_1-k_2)^2 (k_1)^2 (k_1-l)^2 l^2 } \Bigr] \non
&=& \frac{1}{2} \Phi^{(1),T}_{\rho, c} \otimes G^{(0)}_{a,TL,23}(x_1; x_2),
\label{eq:nloctl23}
\eeq }
with the NLO twist-2 transverse wave functions
{\small
\beq
\Phi^{(1),T}_{\rho, a} &=& \frac{- i g^2_s C_F}{8} \mathrm{Tr} \Bigl [
\frac{\gamma^a_{\bot} \gamma^{-}\gamma^a_{\bot} \gamma^{-}(\psl_1 - \ksl_1)
    \gamma^{\rho'} (\psl_1 - \ksl_1 + \lsl) \gamma_{\rho'}}{(p_1 -k_1)^2 (p_1 - k_1 + l)^2 l^2} \Bigr], \non
\Phi^{(1),T}_{\rho, b} &=& \frac{i g^2_s C_F}{8} \mathrm{Tr} \Bigl [
\frac{\gamma^a_{\bot}\gamma^{-} \gamma^{\rho'} (\ksl_1 - \lsl)
    \gamma_{\bot,a} \gamma^{+}(\psl_1 - \ksl_1 + \lsl) \gamma_{\rho'}}{(k_1 - l)^2 (p_1 - k_1 + l)^2 l^2} \Bigr], \non
\Phi^{(1),T}_{\rho, c} &=& \frac{- i g^2_s C_F}{8} \mathrm{Tr}\Bigl[
\frac{ \gamma^a_{\bot} \gamma^{-}\gamma^{\rho'}
    (\ksl_1 - \lsl) \gamma_{\rho'} \ksl_1\gamma^a_{\bot} \gamma^{-} }{(k_1 - l)^2 (k_1 )^2 l^2}\Bigr].
\label{eq:nloabcrhoT}
\eeq }

The NLO infrared amplitudes $G^{(1)}_{X,TL,23}$ for other eight sub-diagrams Figs.~\ref{fig:fig2}(d-k) are of the form of
{\small
\beq
G^{(1)}_{2d,TL,23} &=& e g^4_s \mathrm{Tr} \Bigl [
    \frac{(\psl_1 - \ksl_1 + \lsl)\gamma^{\gamma}
    F_{\alpha \beta \gamma}[\epsl_{1T}\psl_1 \phi^T_{\rho}(x_1)] \gamma^{\alpha} [M_{\rho} \phi^s_{\rho}(x_2)] \gamma_{\beta} (\psl_2 - \ksl_1+ \lsl) \gamma^{\mu} }{(p_2-k_1+l)^2 (k_1-k_2)^2
    (p_1-k_1+l)^2 (k_1-k_2-l)^2 l^2} \Bigr ] \non
 & \sim &\frac{9}{16} \Phi^{(1),T}_{\rho, d}(x_1) \otimes [G^{(0)}_{a,TL,23}(x_1;\xi_1; x_2) - G^{(0)}_{a,TL,23}(\xi_1; x_2)],
\label{eq:nlodtl23}\\
G^{(1)}_{2e,TL,23} &=& - e g^4_s \mathrm{Tr} \Bigl [
    \frac{(\psl_2 - \ksl_1)\gamma^{\mu} F_{\alpha \beta \gamma}[\epsl_{1T}\psl_1 \phi^T_{\rho}(x_1)] \gamma^{\gamma} (\ksl_1 - \lsl) \gamma^{\alpha} [M_{\rho} \phi^s_{\rho}(x_2)] \gamma_{\beta} }{(p_2-k_1)^2 (k_1-k_2)^2 (k_1-l)^2
    (k_1-k_2-l)^2 l^2} \Bigr ] \non
 & \sim & \frac{9}{16} \Phi^{(1),T}_{\rho, e}(x_1) \otimes [G^{(0)}_{a,TL,23}(x_1; x_2) - G^{'(0)}_{a,TL,23}(x_1;\xi_1;x_2)],
\label{eq:nloetl23}\\
G^{(1)}_{2f,TL,23}& = & \frac{ e g^4_s }{9} \mathrm{Tr} \Bigl [
\frac{(\psl_2-\ksl_1+\lsl) \gamma_{\mu} (\psl_1-\ksl_1+\lsl) \gamma_{\rho'} [\epsl_{1T}\psl_1 \phi^T_{\rho}(x_1)]\gamma^{\alpha} [M_{\rho} \phi^s_{\rho}(x_2)] \gamma_{\alpha} (\psl_2 - \ksl_1) \gamma^{\rho'}}{(p_2-k_1)^2 (k_1-k_2)^2 (p_1-k_1+l)^2
(p_2-k_1+l)^2 l^2} \Bigr ]\non
&  \sim &\Phi^{(1),T}_{\rho, d}(x_1) \otimes [G^{(0)}_{a,TL,23}(x_1; x_2) - G^{(0)}_{a,TL,23}(x_1;\xi_1;x_2)] ,
\label{eq:nloftl23}\\
G^{(1)}_{2g,TL,23} &= & \frac{- e g^4_s}{9}  \mathrm{Tr} \Bigl[
\frac{(\psl_2 - \ksl_1+\lsl)\gamma^{\rho'} (\psl_2-\ksl_1) \gamma_{\mu}[\epsl_{1T}\psl_1 \phi^T_{\rho}(x_1)]\gamma_{\rho'}
(\ksl_1 - \lsl) \gamma^{\alpha} [M_{\rho} \phi^s_{\rho}(x_2)] \gamma_{\alpha} }{(p_2-k_1)^2 (k_1-k_2-l)^2
(p_2-k_1+l)^2 (k_1-l)^2 l^2} \Bigr] \non
 & \sim&  \Phi^{(1),T}_{\rho, e}(x_1) \otimes [G^{'(0)}_{a,TL,23}(x_1;\xi_1; x_2) - G^{(0)}_{a,TL,23}(\xi_1;x_2)] .
\label{eq:nlogtl23}\\
G^{(1)}_{2h,TL,23} &= & \frac{e g^4_s }{9} \mathrm{Tr} \Big[
    \frac{(\psl_2-\ksl_1+\lsl)\gamma^{\mu} (\psl_1-\ksl_1+\lsl) \gamma_{\rho'}[\epsl_{1T}\psl_1 \phi^T_{\rho}(x_1)]\gamma^{\alpha} [M_{\rho} \phi^s_{\rho}(x_2)] \gamma^{\rho'} (\psl_2-\ksl_2+\lsl) \gamma_{\alpha} }{(k_1-k_2)^2 (p_2-k_1+l)^2 (p_1-k_1+l)^2
    (p_2-k_2+l)^2 l^2} \Bigr] \non
&\sim & -\frac{1}{8} \Phi^{(1),T}_{\rho, e}(x_1) \otimes G^{(0)}_{a,TL,23}(x_1,\xi_1;x_2),
\label{eq:nlohtl23}\\
G^{(1)}_{2i,TL,23} &= & \frac{- e g^4_s }{9} \mathrm{Tr} \Bigl[
    \frac{(\psl_2-\ksl_1+\lsl)\gamma_{\mu} (\psl_1-\ksl_1+\lsl) \gamma_{\rho'}[\epsl_{1T}\psl_1 \phi^T_{\rho}(x_1)]\gamma^{\alpha} (\ksl_2+\lsl) \gamma^{\rho'} [M_{\rho} \phi^s_{\rho}(x_2)] \gamma_{\alpha} }{(k_1-k_2-l)^2 (p_2-k_1+l)^2 (p_1-k_1+l)^2
    (k_2+l)^2 l^2} \Bigr]\non
&\sim & \frac{1}{8} \Phi^{(1),T}_{\rho, d}(x_1) \otimes G^{(0)}_{a,TL,23}(\xi_1; x_2).
\label{eq:nloitl23}\\
G^{(1)}_{2j,TL,23}& =& \frac{e g^4_s}{9} \mathrm{Tr} \Bigl[
\frac{\gamma_{\alpha} (\psl_2-\ksl_1) \gamma_{ \mu}[\epsl_{1T}\psl_1 \phi^T_{\rho}(x_1)]\gamma_{\rho'} (\ksl_1-\lsl) \gamma^{\alpha}
(\ksl_2-\lsl) \gamma^{\rho'}[M_{\rho} \phi^s_{\rho}(x_2)]}{(k_1-k_2)^2 (p_2-k_1)^2 (k_1-l)^2 (k_2-l)^2 l^2} \Bigr] \non
&\sim & -\frac{1}{8} \Phi^{(1),T}_{\rho, e}(x_1) \otimes G^{(0)}_{a,TL,23}(x_1; x_2) ,
\label{eq:nlojtl23}\\
G^{(1)}_{2k,TL,23} &= & \frac{ - e g^4_s }{9} \mathrm{Tr} \Big[
\frac{(\psl_2-\ksl_2-\lsl)\gamma_{\alpha}(\psl_2-\ksl_1) \gamma_{\mu}[\epsl_{1T}\psl_1 \phi^T_{\rho}(x_1)]\gamma_{\rho'} (\ksl_1-\lsl) \gamma^{\alpha}[M_{\rho} \phi^s_{\rho}(x_2)]\gamma^{\rho'}}{(k_1-k_2-l)^2
(p_2-k_1)^2 (k_1-l)^2 (p_2-k_2-l)^2 l^2}\Bigr] \non
&\sim & \frac{1}{8} \Phi^{(1),T}_{\rho, e}(x_1) \otimes G^{'(0)}_{a,TL,23}(x_1;\xi_1;x_2),
\label{eq:nloktl23}
\eeq }
with the NLO  twist-2 transverse $\rho$ meson wave functions,
{\small
\beq
\Phi^{(1),T}_{\rho, d}(x_1) &=& \frac{- i g^2_s C_F}{8} \mathrm{Tr}\Big[
    \frac{\gamma^a_{\bot} \gamma^{-}\gamma^a_{\bot} \gamma^{-} (\psl_1 - \ksl_1 + \lsl) \gamma^{\rho} v_{\rho}}
    {(p_1 - k_1 + l)^2 l^2 (v \cdot l)} \Bigr ], \label{eq:nlorhoT-d} \non
\Phi^{(1),T}_{\rho, e}(x_1) &=& \frac{ i g^2_s C_F}{8} \mathrm{Tr} \Big[
    \frac{\gamma^a_{\bot} \gamma^{-}\gamma^{\rho}( \ksl_1 - \lsl)\gamma^a_{\bot} \gamma^{-}v_{\rho}}
    {( k_1 - l)^2 l^2 (v \cdot l)} \Bigr ],\label{eq:nlorhoT-e}
\eeq }
which could be written as a nonlocal matrix element with the Wilson lines.

The newly defined LO hard amplitudes $G^{(0)}_{a,TL,23}$, $G^{(0)}_{a,TL,23}$, and  $G^{'(0)}_{a,TL,23}$, as shown
in Eqs.~(A15-A17,A19-A22,A24-A25), are of the following form
{\small
\beq
G^{(0)}_{a,TL,23}(\xi_1; x_2) &=& \frac{-i e g^2_s C_F}{2} \mathrm{Tr}\Bigl [
\frac{[\epsl_{1T}\psl_1 \phi^T_{\rho}(x_1)]
\gamma^{\alpha} [M_{\rho} \phi^s_{\rho}(x_2)] \gamma_{\alpha} (\psl_2 - \ksl_1+\lsl) \gamma_{\mu}}{(p_2-k_1+l)^2 (k_1-k_2-l)^2}
\Bigr ],
\label{eq:loatl231}\\
G^{(0)}_{a,TL,23}(x_1;\xi_1; x_2) &=& \frac{-i e g^2_s C_F}{2} \mathrm{Tr} \Bigl[
\frac{[\epsl_{1T}\psl_1 \phi^T_{\rho}(x_1)]
\gamma^{\alpha} [M_{\rho} \phi^s_{\rho}(x_2)] \gamma_{\alpha} (\psl_2 - \ksl_1+\lsl) \gamma_{\mu}}{(p_2-k_1+l)^2 (k_1-k_2)^2}
\Bigr ],
\label{eq:loatl232}\\
G^{'(0)}_{a,TL,23}(x_1;\xi_1; x_2) &=& \frac{-i e g^2_s C_F}{2} \mathrm{Tr}\Big[
\frac{[\epsl_{1T}\psl_1 \phi^T_{\rho}(x_1)]
\gamma^{\alpha} [M_{\rho} \phi^s_{\rho}(x_2)] \gamma_{\alpha} (\psl_2 - \ksl_1) \gamma_{\mu}}{(p_2-k_1)^2 (k_1-k_2-l)^2} \Bigr].
\label{eq:loatl233}
\eeq }

\subsection{The NLO amplitudes for $G^{(0)}_{a,TL,32}$}

The eleven NLO amplitudes $G^{(1)}_{X,TL,32}$ for sub-diagrams Figs.~\ref{fig:fig2}(a-k) are the following
{\small
\beq
G^{(1)}_{2a,TL,32} &=& -\frac{1}{2} \frac{e g^4_s C^2_F}{2} \mathrm{Tr} \Bigl[
    \frac{[\epsl_{1T} M_{\rho} \phi^{v}_{\rho}(x_1)+ M_{\rho} i \epsilon_{\mu'\nu\rho\sigma}\gamma^{\mu'} \gamma_5  \epsilon^{\nu}_{1T} n^{\rho} v^{\sigma} \phi^{a}_{\rho}(x_1)]}{(p_2-k_1)^2 (k_1-k_2)^2 (p_1-k_1)^2 (p_1-k_1+l)^2 l^2} \non
    && \cdot \gamma^{\alpha}[\epsl_{2L} M_{\rho} \phi_{\rho}(x_2)]\gamma^{\alpha}(\psl_2 - \ksl_1)\gamma_{\mu}(\psl_1 - \ksl_1)\gamma^{\rho'}
    (\psl_1 - \ksl_1 + \lsl) \gamma_{\rho'} \Bigr]\non
&=& \frac{1}{2} \Phi^{(1),v}_{\rho, a}(x_1) \otimes G^{(0),v}_{a,TL,32}(x_1; x_2)+\frac{1}{2} \Phi^{(1),a}_{\rho, a} \otimes G^{(0),a}_{a,TL,32}(x_1; x_2),
\label{eq:nloatl32}\\
G^{(1)}_{2b,TL,32} &=& \frac{ e g^4_s C^2_F}{2} \mathrm{Tr} \Bigl[
    \frac{[\epsl_{1T} M_{\rho} \phi^{v}_{\rho}(x_1)+ M_{\rho} i \epsilon_{\mu'\nu\rho\sigma}\gamma^{\mu'} \gamma_5  \epsilon^{\nu}_{1T} n^{\rho} v^{\sigma} \phi^{a}_{\rho}(x_1)]}{(p_2-k_1+l)^2 (k_1-k_2-l)^2 (p_1-k_1+l)^2 (k_1-l)^2 l^2} \non
&& \cdot\gamma^{\rho'} (\ksl_1 - \lsl) \gamma^{\alpha} [\epsl_{2L} M_{\rho} \phi_{\rho}(x_2)] \gamma_{\alpha}
    (\psl_2 - \ksl_1 + \lsl)\gamma_{\mu}(\psl_1 - \ksl_1 + \lsl) \gamma^{\rho'}\Bigr]\non
&=& \Phi^{(1),v}_{\rho, b}(x_1) \otimes G^{(0),v}_{a,TL,32}(\xi_1; x_2)+\Phi^{(1),a}_{\rho, b} \otimes G^{(0),a}_{a,TL,32}(\xi_1; x_2),
\label{eq:nlobtl32}\\
G^{(1)}_{2c,TL,32} &=&- \frac{1}{2} \frac{e g^4_s C^2_F}{2} \mathrm{Tr}\Bigl[
    \frac{[\epsl_{1T} M_{\rho} \phi^{v}_{\rho}(x_1)+ M_{\rho} i \epsilon_{\mu'\nu\rho\sigma}\gamma^{\mu'} \gamma_5  \epsilon^{\nu}_{1T} n^{\rho} v^{\sigma} \phi^{a}_{\rho}(x_1)]}{(p_2-k_1)^2 (k_1-k_2)^2 (k_1-l)^2 (k_1)^2 l^2} \non
  &&  \cdot \gamma^{\rho'} (\ksl_1-\lsl)\gamma_{\rho'}\ksl_1\gamma^{\alpha}[\epsl_{2L} M_{\rho} \phi_{\rho}(x_2)] \gamma_{\alpha}
  (\psl_2-\ksl_1)\gamma_{\mu} \Bigr] \non
&=& \frac{1}{2}\Phi^{(1),v}_{\rho, c}(x_1)\otimes G^{(0),v}_{a,TL,32}(x_1; x_2) + \frac{1}{2}\Phi^{(1),a}_{\rho, c}\otimes G^{(0),a}_{a,TL,32}(x_1; x_2) ,
\label{eq:anloctl32}\\
G^{(1)}_{2d,TL,32} &=& e g^4_s \mathrm{Tr}\Bigl[
    \frac{[\epsl_{1T} M_{\rho} \phi^{v}_{\rho}(x_1)+ M_{\rho} i \epsilon_{\mu'\nu\rho\sigma}\gamma^{\mu'} \gamma_5  \epsilon^{\nu}_{1T} n^{\rho} v^{\sigma} \phi^{a}_{\rho}(x_1)]}{(p_2-k_1+l)^2 (k_1-k_2)^2 (p_1-k_1+l)^2 (k_1-k_2-l)^2 l^2} \non
   &&  \cdot \gamma^{\alpha} [\epsl_{2L} M_{\rho} \phi_{\rho}(x_2)] \gamma_{\beta} (\psl_2 - \ksl_1+ \lsl) \gamma^{\mu} (\psl_1 - \ksl_1 + \lsl)\gamma^{\gamma} F_{\alpha \beta \gamma}
   \Bigr]\non
& \sim & 0,
\label{eq:nlodtl32}
\eeq
\beq
G^{(1)}_{2e,TL,32} &= & - e g^4_s \mathrm{Tr} \Bigl[
    \frac{[\epsl_{1T} M_{\rho} \phi^{v}_{\rho}(x_1)+ M_{\rho} i \epsilon_{\mu'\nu\rho\sigma}\gamma^{\mu'} \gamma_5  \epsilon^{\nu}_{1T} n^{\rho} v^{\sigma} \phi^{a}_{\rho}(x_1)] }{(p_2-k_1)^2 (k_1-k_2)^2 (k_1-l)^2 (k_1-k_2-l)^2 l^2} \non
   && \cdot\gamma^{\gamma} (\ksl_1 - \lsl) \gamma^{\alpha} [\epsl_{2L} M_{\rho} \phi_{\rho}(x_2)] \gamma_{\beta} (\psl_2 - \ksl_1)
    \gamma^{\mu} F_{\alpha \beta \gamma} \Bigr] \non
 & \sim & \frac{9}{8} \Phi^{(1),v}_{\rho, e} \otimes [G^{(0),v}_{a,TL,32}(x_1; x_2) - G^{'(0),v}_{a,TL,32}(x_1;\xi_1;x_2)]\non
&& + \frac{9}{8} \Phi^{(1),a}_{\rho, e}(x_1) \otimes [G^{(0),a}_{a,TL,32}(x_1; x_2) - G^{'(0),a}_{a,TL,32}(x_1;\xi_1;x_2)],
\label{eq:nloetl32}\\
G^{(1)}_{2f,TL,32}  &=& \frac{ e g^4_s }{9}  \mathrm{Tr}\Bigl[
\frac{[\epsl_{1T} M_{\rho} \phi^{v}_{\rho}(x_1)+ M_{\rho} i \epsilon_{\mu'\nu\rho\sigma}\gamma^{\mu'} \gamma_5  \epsilon^{\nu}_{1T} n^{\rho} v^{\sigma} \phi^{a}_{\rho}(x_1)]}{(p_2-k_1)^2 (k_1-k_2)^2 (p_1-k_1+l)^2 (p_2-k_1+l)^2 l^2} \non
&&\cdot \gamma^{\alpha} [\epsl_{2L} M_{\rho} \phi_{\rho}(x_2)] \gamma_{\alpha} (\psl_2 - \ksl_1) \gamma^{\rho'}(\psl_2-\ksl_1+\lsl) \gamma_{\mu}
(\psl_1-\ksl_1+\lsl) \gamma_{\rho'}\Bigr]\non
& \sim&   \Phi^{(1),v}_{\rho, d} \otimes [G^{(0),v}_{a,TL,32}(x_1; x_2) - G^{(0),v}_{a,TL,32}(x_1;\xi_1;x_2)]\non
&& + \Phi^{(1),a}_{\rho, d}(x_1) \otimes [G^{(0),a}_{a,TL,32}(x_1; x_2) - G^{(0),a}_{a,TL,32}(x_1;\xi_1;x_2)] ,
\label{eq:nloftl32}\\
G^{(1)}_{2g,TL,32} &= & \frac{- e g^4_s }{9} \mathrm{Tr} \Bigl[
\frac{[\epsl_{1T} M_{\rho} \phi^{v}_{\rho}(x_1)+ M_{\rho} i \epsilon_{\mu'\nu\rho\sigma}\gamma^{\mu'} \gamma_5  \epsilon^{\nu}_{1T} n^{\rho} v^{\sigma} \phi^{a}_{\rho}(x_1)]}{(p_2-k_1)^2 (k_1-k_2-l)^2 (p_2-k_1+l)^2 (k_1-l)^2 l^2} \non
&& \cdot\gamma_{\rho'} (\ksl_1 - \lsl) \gamma^{\alpha} [\epsl_{2L} M_{\rho} \phi_{\rho}(x_2)] \gamma_{\alpha} (\psl_2 - \ksl_1+\lsl)
\gamma^{\rho'} (\psl_2-\ksl_1) \gamma_{\mu}\Bigr]\non
& \sim &   0 .
\label{eq:nlogtl32}\\
G^{(1)}_{2h,TL,32} &=& \frac{e g^4_s }{9}  \mathrm{Tr}\Bigl[
    \frac{[\epsl_{1T} M_{\rho} \phi^{v}_{\rho}(x_1)+ M_{\rho} i \epsilon_{\mu'\nu\rho\sigma}\gamma^{\mu'} \gamma_5  \epsilon^{\nu}_{1T} n^{\rho} v^{\sigma} \phi^{a}_{\rho}(x_1)]}{(k_1-k_2)^2 (p_2-k_1+l)^2 (p_1-k_1+l)^2 (p_2-k_2+l)^2 l^2} \non
&&   \cdot \gamma^{\alpha} [\epsl_{2L} M_{\rho} \phi_{\rho}(x_2)] \gamma^{\rho'} (\psl_2-\ksl_2+\lsl) \gamma_{\alpha} (\psl_2-\ksl_1+\lsl)\gamma^{\mu}
   (\psl_1-\ksl_1+\lsl) \gamma_{\rho'}\Bigr]\non
&\sim & 0,
\label{eq:nlohtl32}\\
G^{(1)}_{2i,TL,32} &=& \frac{-e g^4_s }{9} \mathrm{Tr}\Bigl[
    \frac{[\epsl_{1T} M_{\rho} \phi^{v}_{\rho}(x_1)+ M_{\rho} i \epsilon_{\mu'\nu\rho\sigma}\gamma^{\mu'} \gamma_5  \epsilon^{\nu}_{1T} n^{\rho} v^{\sigma} \phi^{a}_{\rho}(x_1)]}{(k_1-k_2-l)^2 (p_2-k_1+l)^2 (p_1-k_1+l)^2 (k_2+l)^2 l^2} \non
&&    \cdot\gamma^{\alpha} (\ksl_2+\lsl) \gamma^{\rho'} [\epsl_{2L} M_{\rho} \phi_{\rho}(x_2)] \gamma_{\alpha} (\psl_2-\ksl_1+\lsl)\gamma_{\mu} (\psl_1-\ksl_1+\lsl) \gamma_{\rho'}
    \Bigr]\non
& \sim&  0.
\label{eq:nloitl32}\\
G^{(1)}_{2j,TL,32} &= & \frac{e g^4_s }{9} \mathrm{Tr}\Bigl[
\frac{[\epsl_{1T} M_{\rho} \phi^{v}_{\rho}(x_1)+ M_{\rho} i \epsilon_{\mu'\nu\rho\sigma}\gamma^{\mu'} \gamma_5  \epsilon^{\nu}_{1T} n^{\rho} v^{\sigma} \phi^{a}_{\rho}(x_1)]}{(k_1-k_2)^2 (p_2-k_1)^2 (k_1-l)^2 (k_2-l)^2 l^2} \non
&& \cdot\gamma_{\rho'} (\ksl_1-\lsl) \gamma^{\alpha} (\ksl_2-\lsl) \gamma^{\rho'}
[\epsl_{2L} M_{\rho} \phi_{\rho}(x_2)]\gamma_{\alpha} (\psl_2-\ksl_1) \gamma_{ \mu} \Bigr]\non
& \sim&  -\frac{1}{8} \Phi^{(1),v}_{\rho, d}(x_1) \otimes G^{(0),v}_{a,TL,32}(x_1; x_2)
        -\frac{1}{8} \Phi^{(1),a}_{\rho, d}(x_1) \otimes G^{(0),a}_{a,TL,32}(x_1; x_2),
\label{eq:nlojtl32}\\
G^{(1)}_{2k,TL,32} &= & \frac{ - e g^4_s }{9} \mathrm{Tr}\Bigl[
\frac{[\epsl_{1T} M_{\rho} \phi^{v}_{\rho}(x_1)+ M_{\rho} i \epsilon_{\mu'\nu\rho\sigma}\gamma^{\mu'} \gamma_5  \epsilon^{\nu}_{1T} n^{\rho} v^{\sigma} \phi^{a}_{\rho}(x_1)]}{(k_1-k_2-l)^2 (p_2-k_1)^2 (k_1-l)^2 (p_2-k_2-l)^2 l^2} \non
&& \cdot\gamma_{\rho'} (\ksl_1-\lsl) \gamma^{\alpha}[\epsl_{2L} M_{\rho} \phi_{\rho}(x_2)]\gamma^{\rho'}(\psl_2-\ksl_2-\lsl)
 \gamma_{\alpha}(\psl_2-\ksl_1) \gamma_{\mu}\Bigr]\non
& \sim& \frac{1}{8} \Phi^{(1),v}_{\rho, e}(x_1) \otimes G^{'(0),v}_{a,TL,32}(x_1;\xi_1; x_2)
      +\frac{1}{8} \Phi^{(1),a}_{\rho, e}(x_1) \otimes G^{'(0),a}_{a,TL,32}(x_1;\xi_1;x_2) ,
\label{eq:nloktl32}
\eeq }
where the NLO twist-3 transverse $\rho$ meson wave functions are the following
{\small
\beq
\Phi^{(1),v}_{\rho, a}(x_1) &=& \frac{- i g^2_s C_F}{4} \mathrm{Tr}\Bigl[  \frac{[\gamma_{\bot\rho} \gamma^{\bot}_{\rho}]
(\psl_1-\ksl_1) \gamma^{\rho'} (\psl_1-\ksl_1+\lsl)\gamma_{\rho'}}{(p_1-k_1)^2 (p_1-k_1+l)^2 l^2}\Bigr], \non
\Phi^{(1),a}_{\rho, a}(x_1) &=& \frac{- i g^2_s C_F}{4} \mathrm{Tr}\Bigl[  \frac{[\gamma_{\bot\rho} \gamma_5]
[\gamma_5 \gamma^{\rho}_{\bot}] (\psl_1-\ksl_1) \gamma^{\rho'} (\psl_1-\ksl_1+\lsl)
\gamma_{\rho'}}{(p_1-k_1)^2 (p_1-k_1+l)^2 l^2} \Bigr],
\label{eq:phi-10a}\\
\Phi^{(1),v}_{\rho, b}(x_1) &=& \frac{i g^2_s C_F}{4} \mathrm{Tr}\Bigl[
\frac{\gamma_{\bot\rho}\gamma^{\rho'}(\ksl_1-\lsl) \gamma_{\bot\rho}
(\psl_1-\ksl_1+\lsl)\gamma_{\rho'}} {(p_1-k_1+l)^2 (k_1-l)^2 l^2} \Bigr], \non
\Phi^{(1),a}_{\rho, b}(x_1) &=& \frac{i g^2_s C_F}{4} \mathrm{Tr} \Bigl[
\frac{[\gamma_{\bot\rho}\gamma_5]\gamma^{\rho'}(\ksl_1-\lsl)
[\gamma_5\gamma^{\rho}_{\bot}  (\psl_1-\ksl_1+\lsl)\gamma_{\rho'}}{(p_1-k_1+l)^2 (k_1-l)^2 l^2}\Bigr],
\label{eq:phi-10b}\\
\Phi^{(1),v}_{\rho, c}(x_1) &=& \frac{- i g^2_s C_F}{4} \mathrm{Tr} \Bigl[
\frac{ \gamma_{\bot\rho} \gamma^{\rho'} (\ksl_1-\lsl)\gamma_{\rho'}\ksl_1 \gamma^{\rho}_{\bot}}
{(k_1-l)^2 (k_1)^2 l^2} \Bigr],\non
\Phi^{(1),a}_{\rho, c}(x_1) &=& \frac{- i g^2_s C_F}{4} \mathrm{Tr} \Bigl[
 \frac{ [\gamma_{\bot\rho}\gamma_5] \gamma^{\rho'} (\ksl_1-\lsl)\gamma_{\rho'}\ksl_1 \gamma_5\gamma^{\rho}_{\bot}}{(k_1-l)^2 (k_1)^2 l^2}
 \Bigr] .
\label{eq:nloabcrhova}\\
\Phi^{(1),v}_{\rho, d}(x_1) &=& \frac{- i g^2_s C_F}{4} \mathrm{Tr} \Bigl[
    \frac{[\gamma_{\bot\rho} \gamma^{\bot}_{\rho}] (\psl_1-\ksl_1+\lsl) \gamma^{\rho}  v_{\rho'}}
    {(p_1-k_1+l)^2 l^2 (v\cdot l)} \Bigr],\non
\Phi^{(1),a}_{\rho, d}(x_1) &=& \frac{- i g^2_s C_F}{4} \mathrm{Tr}\Bigl[
    \frac{[\gamma_{\bot\rho} \gamma_5][\gamma_5 \gamma^{\rho}_{\bot}](\psl_1-\ksl_1+\lsl) \gamma^{\rho}  v_{\rho'}}
    {(p_1-k_1+l)^2 l^2 (v\cdot l)} \Bigr].
\label{eq:nlorhova-d}\\
\Phi^{(1),v}_{\rho, e}(x_1) &&= \frac{i g^2_s C_F}{4} \mathrm{Tr}\Bigl[
    \frac{[\gamma_{\bot\rho}] \gamma^{\rho} (\ksl_1-\lsl) [\gamma^{\rho}_{\bot}] v_{\rho'}}
    {(k_1-l)^2 l^2 (v\cdot l)} \Bigr],\non
\Phi^{(1),a}_{\rho, e}(x_1) &&= \frac{i g^2_s C_F}{4} \mathrm{Tr}\Bigl [
    \frac{[\gamma_{\bot\rho}\gamma_5] \gamma^{\rho} (\ksl_1-\lsl) [\gamma_5\gamma^{\rho}_{\bot}] v_{\rho'}}
    {(k_1-l)^2 l^2 (v\cdot l)} \Bigr] .
\label{eq:nlorhova-e}
\eeq}
The new LO hard amplitudes $G^{(0)}_{a,TL,32}$ and $G^{'(0)}_{a,TL,32}$ appeared in Eqs.~(A31-A41 ) are the following
{\small
\beq
G^{(0)}_{a,TL,32}(\xi_1;x_2) &=& -\frac{i e g^2_s C_F}{2} \mathrm{Tr}\Bigl[
\frac{[\epsl_{1T} M_{\rho} \phi^{v}_{\rho}(x_1)
+ M_{\rho} i \epsilon_{\mu'\nu\rho\sigma}\gamma^{\mu'} \gamma_5  \epsilon^{\nu}_{1T} n^{\rho} v^{\sigma} \phi^{a}_{\rho}(x_1)]
}{(p_2-k_1+l)^2 (k_1-k_2-l)^2}\non
&&\cdot \gamma^{\alpha}[\epsl_{2L} M_{\rho} \phi_{\rho}(x_2)]  \gamma_{\alpha} (\psl_2 - \ksl_1 +\lsl) \gamma_{\mu}\Bigr] ,
\label{eq:nlotl321}\\
G^{(0)}_{a,TL,32}(x_1,\xi_1;x_2) &=& -\frac{i e g^2_s C_F}{2} \mathrm{Tr} \Bigl[
\frac{[\epsl_{1T} M_{\rho} \phi^{v}_{\rho}(x_1)
+ M_{\rho} i \epsilon_{\mu'\nu\rho\sigma}\gamma^{\mu'} \gamma_5  \epsilon^{\nu}_{1T} n^{\rho} v^{\sigma} \phi^{a}_{\rho}(x_1)]
}{(p_2-k_1+l)^2 (k_1-k_2)^2}\non
&&\cdot \gamma^{\alpha}[\epsl_{2L} M_{\rho} \phi_{\rho}(x_2)]  \gamma_{\alpha} (\psl_2 - \ksl_1 +\lsl) \gamma_{\mu} \Bigr],
\label{eq:nlotl322}\\
G^{'(0)}_{a,TL,32}(x_1,\xi_1;x_2) &=& -\frac{i e g^2_s C_F}{2} \mathrm{Tr} \Bigl[
\frac{[\epsl_{1T} M_{\rho} \phi^{v}_{\rho}(x_1)
+ M_{\rho} i \epsilon_{\mu'\nu\rho\sigma}\gamma^{\mu'} \gamma_5  \epsilon^{\nu}_{1T} n^{\rho} v^{\sigma} \phi^{a}_{\rho}(x_1)]
}{(p_2-k_1)^2 (k_1-k_2-l)^2}\non
&&\cdot \gamma^{\alpha}[\epsl_{2L} M_{\rho} \phi_{\rho}(x_2)]  \gamma_{\alpha} (\psl_2 - \ksl_1 ) \gamma_{\mu}\Bigr].
\label{eq:nlotl323}
\eeq }

\subsection{The NLO amplitudes for $G^{(0)}_{a,TT,33}$}

The eleven NLO amplitudes $G^{(1)}_{X,TT,32}$ for sub-diagrams Figs.~\ref{fig:fig2}(a-k) are the following
{\small
\beq
G^{(1)}_{2a,TT,33} &=& -\frac{1}{2} \frac{e g^4_s C^2_F}{2}  \mathrm{Tr} \Bigl[
    \frac{ (\psl_1 - \ksl_1 + \lsl) \gamma_{\rho'}[\epsl_{1T} M_{\rho} \phi^{v}_{\rho}(x_1)+M_{\rho} i \epsilon_{\mu'\nu\rho\sigma}\gamma^{\mu'} \gamma_5  \epsilon^{\nu}_{1T} n^{\rho} v^{\sigma} \phi^{a}_{\rho}(x_1)]
    \gamma^{\alpha}}{(p_2-k_1)^2 (k_1-k_2)^2 (p_1-k_1)^2 (p_1-k_1+l)^2 l^2} \non
 &&   \cdot[\epsl_{2T} M_{\rho} \phi^{v}_{\rho}(x_2)+ M_{\rho} i \epsilon_{\mu'\nu\rho\sigma} \gamma_5 \gamma^{\mu'} \epsilon^{\nu}_{2T} v^{\rho} n^{\sigma} \phi^{a}_{\rho}(x_2)]\gamma^{\alpha}
    (\psl_2 - \ksl_1)\gamma_{\mu}(\psl_1 - \ksl_1)\gamma^{\rho'} \Bigr] \non
&=& \frac{1}{2} \Phi^{(1),v}_{\rho, a}(x_1) \otimes G^{(0),v}_{a,TT,33}(x_1; x_2)+\phi^{(1),a}_{\rho, a} \otimes G^{(0),a}_{a,TT,33}(x_1; x_2),
\label{eq:4a}\\
G^{(1)}_{2b,TT,33} &=& \frac{ e g^4_s C^2_F}{2} \mathrm{Tr}\Bigl[
    \frac{\gamma_{\mu}(\psl_1 - \ksl_1 + \lsl) \gamma^{\rho'}[\epsl_{1T} M_{\rho} \phi^{v}_{\rho}(x_1)+M_{\rho} i \epsilon_{\mu'\nu\rho\sigma}\gamma^{\mu'} \gamma_5  \epsilon^{\nu}_{1T} n^{\rho} v^{\sigma} \phi^{a}_{\rho}(x_1)]}
    {(p_2-k_1+l)^2 (k_1-k_2-l)^2 (p_1-k_1+l)^2 (k_1-l)^2 l^2} \non
&&    \cdot\gamma^{\rho'} (\ksl_1 - \lsl) \gamma^{\alpha} [\epsl_{2T} M_{\rho} \phi^{v}_{\rho}(x_2)+ M_{\rho} i \epsilon_{\mu'\nu\rho\sigma} \gamma_5 \gamma^{\mu'} \epsilon^{\nu}_{2T} v^{\rho} n^{\sigma} \phi^{a}_{\rho}(x_2)]
\gamma_{\alpha}(\psl_2 - \ksl_1 + \lsl) \Bigr]\non
&=& \Phi^{(1),v}_{\rho, b}(x_1) \otimes G^{(0),v}_{a,TT,33}(\xi_1, x_2)+\Phi^{(1),a}_{\rho, b} \otimes G^{(0),a}_{a,TT,33}(\xi_1, x_2),
\label{eq:4b}\\
G^{(1)}_{2c,TT,33} &=&- \frac{1}{2} \frac{e g^4_s C^2_F}{2} \mathrm{Tr}\Bigl[
    \frac{\gamma_{\alpha} (\psl_2-\ksl_1)\gamma_{\mu}[\epsl_{1T} M_{\rho} \phi^{v}_{\rho}(x_1)
+M_{\rho} i \epsilon_{\mu'\nu\rho\sigma}\gamma^{\mu'} \gamma_5  \epsilon^{\nu}_{1T} n^{\rho}
v^{\sigma} \phi^{a}_{\rho}(x_1)]} {(p_2-k_1)^2 (k_1-k_2)^2 (k_1-l)^2 (k_1)^2 l^2} \non
&&    \cdot \gamma^{\rho'} (\ksl_1-\lsl)\gamma_{\rho'}\ksl_1\gamma^{\alpha}[\epsl_{2T}
M_{\rho} \phi^{v}_{\rho}(x_2)+ M_{\rho} i \epsilon_{\mu'\nu\rho\sigma} \gamma_5 \gamma^{\mu'}
\epsilon^{\nu}_{2T} v^{\rho} n^{\sigma} \phi^{a}_{\rho}(x_2)] \Bigr]\non
&=& \frac{1}{2}\Phi^{(1),v}_{\rho, c}(x_1)\otimes G^{(0),v}_{a,TT,33}(x_1; x_2) +\frac{1}{2}
\Phi^{(1),a}_{\rho, c}(x_1)\otimes G^{(0),a}_{a,TT,33}(x_1; x_2) ,
\label{eq:4c}\\
G^{(1)}_{2d,TT,33} &=&- e g^4_s   \mathrm{Tr} \Bigl[
    \frac{(\psl_1 - \ksl_1 + \lsl)\gamma^{\gamma} F_{\alpha \beta \gamma}
[\epsl_{1T} M_{\rho} \phi^{v}_{\rho}(x_1)+M_{\rho} i \epsilon_{\mu'\nu\rho\sigma}\gamma^{\mu'} \gamma_5
\epsilon^{\nu}_{1T} n^{\rho} v^{\sigma} \phi^{a}_{\rho}(x_1)]}{(p_2-k_1+l)^2 (k_1-k_2)^2 (p_1-k_1+l)^2
(k_1-k_2-l)^2 l^2} \non
&&     \cdot\gamma^{\alpha} [\epsl_{2T} M_{\rho} \phi^{v}_{\rho}(x_2)+ M_{\rho} i \epsilon_{\mu'\nu\rho\sigma}
\gamma_5 \gamma^{\mu'} \epsilon^{\nu}_{2T} v^{\rho} n^{\sigma} \phi^{a}_{\rho}(x_2)]\gamma_{\beta}
(\psl_2 - \ksl_1+ \lsl) \gamma^{\mu} \Bigr ]\non
& \sim &  0,
\label{eq:4d}\\
G^{(1)}_{2e,TT,33} &= & - e g^4_s  \mathrm{Tr}\Bigl[
    \frac{(\psl_2 - \ksl_1)\gamma^{\mu} F_{\alpha \beta \gamma}
    [\epsl_{1T} M_{\rho} \phi^{v}_{\rho}(x_1)+M_{\rho} i \epsilon_{\mu'\nu\rho\sigma}\gamma^{\mu'}
    \gamma_5  \epsilon^{\nu}_{1T} n^{\rho} v^{\sigma} \phi^{a}
    _{\rho}(x_1)] }
    {(p_2-k_1)^2 (k_1-k_2)^2 (k_1-l)^2 (k_1-k_2-l)^2 l^2} \non
&&   \cdot \gamma^{\gamma} (\ksl_1 - \lsl) \gamma^{\alpha} [\epsl_{2T} M_{\rho} \phi^{v}_{\rho}(x_2)
+ M_{\rho} i \epsilon_{\mu'\nu\rho\sigma} \gamma_5
\gamma^{\mu'} \epsilon^{\nu}_{2T} v^{\rho} n^{\sigma} \phi^{a}_{\rho}(x_2)] \gamma_{\beta} \Bigr] \non
&\sim & \frac{9}{16} \Phi^{(1),v}_{\rho, e} (x_1)\otimes \bigl[G^{(0),v}_{a,TT,33}(x_1; x_2)
- G^{'(0),v}_{a,TT,33}(x_1;\xi_1;x_2)\bigr] \non
&& + \frac{9}{16} \Phi^{(1),a}_{\rho, e}(x_1) \otimes \bigl[G^{(0),a}_{a,TT,33}(x_1; x_2) - G^{'(0),a}_{a,TT,33}(x_1;\xi_1;x_2)\bigr],
\label{eq:4e}\\
G^{(1)}_{2f,TT,33} &= & \frac{e g^4_s}{9} \mathrm{Tr}\Bigl[
\frac{(\psl_2-\ksl_1+\lsl) \gamma_{\mu} (\psl_1-\ksl_1+\lsl)
\gamma_{\rho'}[\epsl_{1T} M_{\rho} \phi^{v}_{\rho}(x_1)+M_{\rho}
i \epsilon_{\mu'\nu\rho\sigma}\gamma^{\mu'} \gamma_5  \epsilon^{\nu}_{1T} n^{\rho} v^{\sigma} \phi^{a}_{\rho}(x_1)]}
{(p_2-k_1)^2 (k_1-k_2)^2 (p_1-k_1+l)^2 (p_2-k_1+l)^2 l^2} \non
&&\cdot \gamma^{\alpha} [\epsl_{2T} M_{\rho} \phi^{v}_{\rho}(x_2)
+ M_{\rho} i \epsilon_{\mu'\nu\rho\sigma} \gamma_5 \gamma^{\mu'}
\epsilon^{\nu}_{2T} v^{\rho} n^{\sigma} \phi^{a}_{\rho}(x_2)] \gamma_{\alpha} (\psl_2 - \ksl_1) \gamma^{\rho'}
\Bigr]\non
&\sim &  0 ,
\label{eq:4f}\\
G^{(1)}_{2g,TT,33} &= & \frac{ - e g^4_s }{9} \mathrm{Tr}\Bigl[
\frac{(\psl_2 - \ksl_1+\lsl)\gamma^{\rho'} (\psl_2-\ksl_1) \gamma_{\mu}[\epsl_{1T}
M_{\rho} \phi^{v}_{\rho}(x_1)+M_{\rho} i \epsilon_{\mu'\nu\rho\sigma}\gamma^{\mu'}
\gamma_5  \epsilon^{\nu}_{1T} n^{\rho} v^{\sigma} \phi^{a}_{\rho}(x_1)]}
{(p_2-k_1)^2 (k_1-k_2-l)^2 (p_2-k_1+l)^2 (k_1-l)^2 l^2} \non
&&\cdot\gamma_{\rho'} (\ksl_1 - \lsl) \gamma^{\alpha} [\epsl_{2T} M_{\rho} \phi^{v}_{\rho}(x_2)
+ M_{\rho} i \epsilon_{\mu'\nu\rho\sigma} \gamma_5 \gamma^{\mu'} \epsilon^{\nu}_{2T}
v^{\rho} n^{\sigma} \phi^{a}_{\rho}(x_2)] \gamma_{\alpha}
\Bigr] \non
&\sim & \Phi^{(1),v}_{\rho, e}(x_1) \otimes [ G^{'(0),v}_{a,TT,33}(x_1;\xi_1;x_2)-G^{(0),v}_{a,TT,33}(\xi_1;x_2)]\non
&&+\Phi^{(1),a}_{\rho, e}(x_1) \otimes[ G^{'(0),a}_{a,TT,33}(x_1;\xi_1;x_2)-G^{(0),a}_{a,TT,33}(\xi_1;x_2)] .
\label{eq:4g}
\eeq
\beq
G^{(1)}_{2h,TT,33} &= & \frac{e g^4_s}{9} \mathrm{Tr}\Bigl[
    \frac{\gamma^{\mu} (\psl_1-\ksl_1+\lsl) \gamma_{\rho'}[\epsl_{1T} M_{\rho} \phi^{v}_{\rho}(x_1)+M_{\rho}
    i \epsilon_{\mu'\nu\rho\sigma}\gamma^{\mu'}
    \gamma_5  \epsilon^{\nu}_{1T} n^{\rho} v^{\sigma} \phi^{a}_{\rho}(x_1)]}
    {(k_1-k_2)^2 (p_2-k_1+l)^2 (p_1-k_1+l)^2 (p_2-k_2+l)^2 l^2} \non
&&    \cdot\gamma^{\alpha} [\epsl_{2T} M_{\rho} \phi^{v}_{\rho}(x_2)
+ M_{\rho} i \epsilon_{\mu'\nu\rho\sigma} \gamma_5 \gamma^{\mu'} \epsilon^{\nu}_{2T}
v^{\rho} n^{\sigma} \phi^{a}_{\rho}(x_2)] \gamma^{\rho'} (\psl_2-\ksl_2+\lsl)\gamma_{\alpha} (\psl_2-\ksl_1+\lsl)\Bigr] \non
&\sim & 0,
\label{eq:4h}\\
G^{(1)}_{2i,TT,33} &= & \frac{-e g^4_s }{9} \mathrm{Tr} \Bigl[
\frac{\gamma_{\mu} (\psl_1-\ksl_1+\lsl) \gamma_{\rho'}[\epsl_{1T} M_{\rho} \phi^{v}_{\rho}(x_1)+M_{\rho} i \epsilon_{\mu'\nu\rho\sigma}\gamma^{\mu'} \gamma_5  \epsilon^{\nu}_{1T} n^{\rho} v^{\sigma} \phi^{a}_{\rho}(x_1)]}
    {(k_1-k_2-l)^2 (p_2-k_1+l)^2 (p_1-k_1+l)^2 (k_2+l)^2 l^2} \non
&&    \cdot\gamma^{\alpha} (\ksl_2+\lsl) \gamma^{\rho'} [\epsl_{2T} M_{\rho} \phi^{v}_{\rho}(x_2)+ M_{\rho} i \epsilon_{\mu'\nu\rho\sigma} \gamma_5 \gamma^{\mu'} \epsilon^{\nu}_{2T} v^{\rho} n^{\sigma} \phi^{a}_{\rho}(x_2)]
\gamma_{\alpha} (\psl_2-\ksl_1+\lsl) \Bigr]\non
&\sim & 0.
\label{eq:4i}\\
G^{(1)}_{2j,TT,33} &= & \frac{e g^4_s }{9} \mathrm{Tr}\Bigl[
\frac{\gamma_{\alpha} (\psl_2-\ksl_1) \gamma_{\mu}[\epsl_{1T}
M_{\rho} \phi^{v}_{\rho}(x_1)+M_{\rho} i \epsilon_{\mu'\nu\rho\sigma}\gamma^{\mu'} \gamma_5
\epsilon^{\nu}_{1T} n^{\rho} v^{\sigma} \phi^{a}_{\rho}(x_1)]}{(k_1-k_2)^2 (p_2-k_1)^2 (k_1-l)^2 (k_2-l)^2 l^2} \non
&&\cdot\gamma_{\rho'} (\ksl_1-\lsl) \gamma^{\alpha} (\ksl_2-\lsl) \gamma^{\rho'}[\epsl_{2T}
M_{\rho} \phi^{v}_{\rho}(x_2)+ M_{\rho} i \epsilon_{\mu'\nu\rho\sigma} \gamma_5 \gamma^{\mu'}
\epsilon^{\nu}_{2T} v^{\rho} n^{\sigma} \phi^{a}_{\rho}(x_2)]\Bigr] \non
&\sim & -\frac{1}{8} \Phi^{(1),v}_{\rho, e}(x_1) \otimes G^{(0),v}_{a,TT,33}(x_1; x_2)
 - \frac{1}{8} \Phi^{(1),a}_{\rho, e}(x_1) \otimes G^{(0),a}_{a,TT,33}(x_1; x_2) ,
\label{eq:4j}\\
G^{(1)}_{2k,TT,33} &= & \frac{- e g^4_s }{9} \mathrm{Tr}\Bigl[
\frac{(\psl_2-\ksl_2-\lsl)\gamma_{\alpha}(\psl_2-\ksl) \gamma_{\mu}[\epsl_{1T} M_{\rho}
\phi^{v}_{\rho}(x_1)+M_{\rho} i \epsilon_{\mu'\nu\rho\sigma}\gamma^{\mu'} \gamma_5
\epsilon^{\nu}_{1T} n^{\rho} v^{\sigma} \phi^{a}_{\rho}(x_1)]}{(k_1-k_2-l)^2 (p_2-k_1)^2 (k_1-l)^2 (p_2-k_2-l)^2 l^2} \non
&&\cdot\gamma_{\rho'} (\ksl_1-\lsl) \gamma^{\alpha}[\epsl_{2T} M_{\rho} \phi^{v}_{\rho}(x_2)
+ M_{\rho} i \epsilon_{\mu'\nu\rho\sigma} \gamma_5 \gamma^{\mu'} \epsilon^{\nu}_{2T}
v^{\rho} n^{\sigma} \phi^{a}_{\rho}(x_2)]\gamma^{\rho'} \Bigr]\non
&\sim &\frac{1}{8} \Phi^{(1),v}_{\rho, e}(x_1) \otimes G^{'(0),v}_{a,TT,33}(x_1;\xi_1; x_2)
 + \frac{1}{8} \Phi^{(1),a}_{\rho, e}(x_1) \otimes G^{'(0),a}_{a,TT,33}(x_1;\xi_1; x_2),
\label{eq:4k}
\eeq }
where the NLO twist-3 transverse $\rho$ meson wave functions  $ (\Phi^{(1),v}_{\rho, i}, \Phi^{(1),a}_{\rho, i} )$
with $i=(a,b,c,e)$  are the same ones as those in Eqs.~(\ref{eq:phi-10a}-\ref{eq:nlorhova-e}).
The new LO hard amplitudes appeared in Eqs.~(\ref{eq:4a}-\ref{eq:4e}) are also defined in a similar way
as those in previous subsection:
{\small
\beq
G^{(0)}_{a,TT,33}(\xi_1;x_2) & =& -\frac{i e g^2_s C_F}{2} \mathrm{Tr} \Bigl[
\frac{[\epsl_{1T} M_{\rho} \phi^{v}_{\rho}(x_1)+M_{\rho} i \epsilon_{\mu'\nu\rho\sigma}\gamma^{\mu'}
\gamma_5  \epsilon^{\nu}_{1T}
n^{\rho} v^{\sigma} \phi^{a}_{\rho}(x_1)]}{(p_2-k_1+l)^2 (k_1-k_2-l)^2}\non
&&\hspace{-1cm}\cdot \gamma^{\alpha} [\epsl_{2T} M_{\rho} \phi^{v}_{\rho}(x_2)+ M_{\rho}
i \epsilon_{\mu'\nu\rho\sigma} \gamma_5 \gamma^{\mu'} \epsilon^{\nu}_{2T}
v^{\rho} n^{\sigma} \phi^{a}_{\rho}(x_2)]  \gamma_{\alpha}(\psl_2 - \ksl_1 +\lsl) \gamma_{\mu}\Bigr],
\label{eq:g0a}\\
G^{(0)}_{a,TT,33}(x_1,\xi_1;x_2) & =& -\frac{i e g^2_s C_F}{2} \mathrm{Tr} \Bigl[
\frac{[\epsl_{1T} M_{\rho} \phi^{v}_{\rho}(x_1)+M_{\rho}
i \epsilon_{\mu'\nu\rho\sigma}\gamma^{\mu'} \gamma_5  \epsilon^{\nu}_{1T}
n^{\rho} v^{\sigma} \phi^{a}_{\rho}(x_1)]}{(p_2-k_1+l)^2 (k_1-k_2)^2}\non
&&\hspace{-1cm}\cdot \gamma^{\alpha} [\epsl_{2T} M_{\rho} \phi^{v}_{\rho}(x_2)
+ M_{\rho} i \epsilon_{\mu'\nu\rho\sigma} \gamma_5 \gamma^{\mu'}
\epsilon^{\nu}_{2T} v^{\rho} n^{\sigma} \phi^{a}_{\rho}(x_2)]  \gamma_{\alpha}(\psl_2 - \ksl_1 +\lsl) \gamma_{\mu}\Bigr],
\label{eq:g0b}\\
G^{'(0)}_{a,TT,33}(x_1,\xi_1;x_2) & =& -\frac{i e g^2_s C_F}{2} \mathrm{Tr} \Bigl[
\frac{[\epsl_{1T} M_{\rho} \phi^{v}_{\rho}(x_1)+M_{\rho} i \epsilon_{\mu'\nu\rho\sigma}
\gamma^{\mu'} \gamma_5  \epsilon^{\nu}_{1T} n^{\rho} v^{\sigma} \phi^{a}_{\rho}(x_1)]}{(p_2-k_1)^2 (k_1-k_2-l)^2}\non
&&\hspace{-1cm}\cdot \gamma^{\alpha} [\epsl_{2T} M_{\rho} \phi^{v}_{\rho}(x_2)+ M_{\rho}
i \epsilon_{\mu'\nu\rho\sigma} \gamma_5 \gamma^{\mu'} \epsilon^{\nu}_{2T}
v^{\rho} n^{\sigma} \phi^{a}_{\rho}(x_2)]  \gamma_{\alpha}(\psl_2 - \ksl_1) \gamma_{\mu}\Bigr] .
\label{eq:g0c}
\eeq}

\end{appendix}

\end{document}